\documentclass[12pt,letterpaper]{article}
\usepackage{osajnl2}
\usepackage{amsmath,amssymb}
\usepackage[hang]{subfigure}
\usepackage{cite}
\usepackage{srcltx}
\usepackage{graphicx}

\usepackage{color}
\definecolor{my-brown}{rgb}{0.69,0.247,0.13}                    %Color = definition of my brown

\def\0#1{{\overset{\scriptscriptstyle \circ}{\mathbf{#1}}{}}}       %over circle: bold-roman
\def\s0#1{{\overset{\scriptscriptstyle \circ}{\boldsymbol{#1}}{}}}  %over circle: bold-symbol
\def\1#1{{\widehat{{\boldsymbol{#1}}}}}                             %widehat: bold-symbol
\def\2#1{\widehat{#1}}                                              %widehat
\def\3#1{{\mathbf{#1}}}                                             %bold-roman
\def\4#1{{\boldsymbol{#1}}}                                         %bold-symbol
\def\5#1{{\cal#1}}                                                  %caligraphy
\def\+#1{{\overset{{\scriptscriptstyle +}}{#1}{}}}                  %superplus
\def\b+#1{{\overset{{\scriptscriptstyle +}}{\mathbf{#1}}{}}}        %superplus  bold-roman
\def\g+#1{{\overset{{\scriptscriptstyle +}}{\boldsymbol{#1}}{}}}    %superplus bold-symbol
\def\6#1{\bar{#1}}                                                  %over-bar
\def\7#1{{\scriptscriptstyle{#1}}}                                  %small
\def\8#1{\widetilde{#1}}
\def\9#1{{\text{#1}}}

\begin{document}

\title{Non-Paraxial Wave Analysis of 3D Airy Beams}
%\author{Yan Kaganovsky and Ehud Heyman}
\author{Yan Kaganovsky,$^{*}$ and Ehud Heyman}
\address{School of Electrical Engineering, Tel Aviv University, \\ Tel-Aviv 69978,  Israel}
\address{$^*$yankagan@gmail.com}

\begin{abstract}
The 3D Airy beam (AiB) is thoroughly explored from a wave-theory point of view. We utilize the exact spectral integral  for the AiB to derive local ray-based solutions that do not suffer from the limitations of the conventional parabolic equation (PE) solution, and are valid far beyond the paraxial zone and for longer ranges. The ray topology near the main lobe of the AiB delineates a hyperbolic umilic diffraction catastrophe, consisting of a cusped double-layered caustic, but this caustic is deformed in the far range where the field loses its beam shape. The field in the vicinity of this caustic is described uniformly by a hyperbolic umilic canonical integral which is structured explicitly on the local geometry of the caustic as obtained from the initial field distribution. In order to accommodate the finite-energy AiB we also modify the canonical integral by adding a complex loss parameter. The canonical integral is calculated using a series expansion and the results are used to identify the validity zone of the conventional PE solution. The analysis is performed within the framework of the non-dispersive AiB where the aperture field is scaled with frequency such that the ray skeleton is frequency-independent. This scaling enables an extension of the theory to the ultra wide band (UWB) regime and ensures that the pulsed field propagates along the curved beam trajectory without dispersion, as will be demonstrated in a subsequent publication.
\end{abstract}

\maketitle

\section{Introduction}
Airy beams (AiB) are exact solutions of the time-harmonic  \emph{parabolic} wave equation (PE) in  free space, that are generated by an Airy function initial condition in an extended aperture. The AiB's have attracted a considerable attention mainly because of their intriguing features, the most distinctive one is the propagation along \emph{curved} trajectories \cite{Sivil_Phyrev, Sivil_Op_let07, bandres07,Sivil_Op_let08}. Other intriguing properties are the \emph{weak diffraction} \cite {Sivil_Op_let07} and \emph{self-healing} (i.e., regeneration of the solutions if the beam is partially obstructed)  \cite{Broky08}. Many interesting applications that utilize these properties have been suggested, e.g. \cite{baumgartl2008, GO_pres_AiB, Gu_2010, Polynkin2009, Arie09}.

In a previous paper \cite{KaganHeyman}, we provided a cogent wave-based description to the 2D-AiB's that clearly explains all these features. Using  uniform asymptotic evaluation of Kirchhoff's radiation integral  \cite{chester}  and  uniform geometrical optics (UGO) \cite{Ludwig65,Kravtsov64}, it has been shown that the AiB \emph{is not} generated by the main lobe of the Airy function in the aperture, but it is in fact a caustic of rays that emerge sideways from the side lobes of the extended aperture distribution. This implies that the AiB cannot be regarded as a ``beam field'' in the sense that the evolution of the main lobe along the curved propagation trajectory is not described by a local wave dynamics. It has also been established that ray theory in conjunction with the UGO provides an accurate solution for the AiB, which fully agrees with the conventional PE solution of \cite{Sivil_Phyrev, Sivil_Op_let07} when the latter is valid. The UGO solution is accurate even at large observation ranges where the PE solution fails due to the curving beam axis. It has been noted that these ray-based tools can be used to construct other curving beam solutions in uniform and non-uniform media by back-projecting the rays from the caustic to the aperture and synthesizing the aperture distribution that focuses onto that caustic \cite{KaganHeyman}. This concept has been applied recently in \cite{AiB_arbitrary_traj_PRL2011}.

Using the ray description of \cite{KaganHeyman} we have also constructed in \cite{KaganHeymanJOSA2011} a class of \emph{non-dispersive} Airy pulsed beams (AiPB). These solutions are obtained by synthesizing the aperture field distribution such that the ray skeleton is the same for all frequencies. This scaling of the aperture distribution, unlike previous suggestions \cite{Saari}, ensures that all  frequency components focus on the same curved caustic, thus giving rise to a pulse that propagates along the curved trajectory with no dispersion. A closed-form time-domain expression for the AiPB has also been derived in \cite{KaganHeymanJOSA2011}  using the spectral theory of transients (STT) \cite{STT1,STT2,STT3}.

The main goal of this paper is to clarify the wave mechanism of the 3D-AiB by deriving uniform ray-based solutions that reveal the local features of the field. Furthermore, these solutions rely on the exact ray and caustic topology of the AiB and therefore  do not suffer from the limitations of the conventional PE solution. We note that due to the curving of the beam trajectory, the PE solution is valid only for short propagation distances and in particular it does not predict  the termination of the  AiB at a certain range, as will be discussed here. In order to meet this goal we first analyze the ray topology of the 3D-AiB (Sec.~\ref{caustic_top}) and then use this analysis to derive uniform solutions that are structured \emph{explicitly} upon the local geometry of the caustic (Secs.~4-5). As will be explained below, we emphasize here \emph{local} uniform solutions rather than \emph{globally} uniform solutions since the former yields explicit solutions that explain all the parameters and features of the field.   The analysis is performed within the framework of the 3D \emph{non-dispersive} AiB, defined in Sec.~\ref{non_dis_AiB} such that the ray skeleton is frequency-independent. The non-dispersive formulation not only simplifies the ray analysis by making it more transparent, but it also enables the construction of the non-dispersive AiPB as noted above. This subject will be explored in a subsequent publication \cite{KaganHeyman_3DAiB_part2_JOSA}, where the results of the present work will be used to synthesize 3D-AiPB and to derive closed-form time-domain solutions.

In Sec.~\ref{caustic_top} we study the exact ray and caustic topology of the 3D-AiB, concentrating on the region near the curved beam axis and deferring  the discussion on the global caustic topology to  Sec.~\ref{caustic_complete}. The analytical details are summarized separately in Appendix~\ref{exact_caustic_eqs}. The topology near the beam axis has many intricate features and it is considerably more complicated than the approximate ray topology of the 3D PE solution in \cite{GO_pres_AiB} or  the 2D case in \cite{KaganHeyman}. Specifically, we show that the main lobe of the AiB is described by a cusped double-layered caustic  which is of the hyperbolic umbilic type \cite{trinkaus1977analysis,berry1980iv,nye1999natural, nye2006dislocation,berry2010axial} (see Figs.~\ref{Fig_caustic_cross_z}--\ref{Fig_caustic_exact}).  We also identify the range where the isolated hyperbolic umbilic  caustic that describes the AiB field near its propagation axis, evolves into the more complicated parabolic umbilic caustic considered in  Sec.~\ref{caustic_complete}, so that the AiB loses its form. It should be noted that these caustic evolutions (unfoldings) are expected from  catastrophe theory principles (see e.g., Fig.~4.4 and Fig.~7.3.(3-5) in \cite{nye1999natural}, respectively), yet, in this paper we provide explicit expressions for this overall topology and also use it to construct the field solution, as described below.

The field calculations are considered next in Secs.~\ref{field_calc}--\ref{catastrophe}. We start from the exact spectral diffraction integral for the AiB, and then discuss its asymptotic evaluation in terms of geometrical optics (GO). The field near the beam axis, where the ray topology is  of the hyperbolic umbilic type, is considered separately in Sec.~\ref{catastrophe} within the framework of diffraction catastrophe theory \cite{poston1996catastrophe,Kravtsov_catastrophe,nye1999natural}, discussed  below.

Catastrophe theory \cite{poston1996catastrophe,Kravtsov_catastrophe,nye1999natural}  classifies the caustic topologies in stable polynomial forms, referred to as elementary  catastrophes. This also provides a systematic framework for calculating the field near caustics with arbitrary topologies. The general methodology is to map the original integral representation of the field, expressed either spatially or spectrally (e.g., Kirchhoff or plane-wave integrals, respectively), onto the diffraction catastrophe integral  corresponding to the elementary catastrophe that describes the caustic. This integral, also referred to as  canonical integral, constitutes a standard special function \cite[Sec.~36.2]{DLMF}.

The mapping of the  diffraction integral into the canonical catastrophe integral may be applied globally via a systematic method first introduced in \cite[Sec.~5]{berry1976waves}. This method extends to multi-dimensions the method of Chester et al. \cite{chester} (see also \cite{bleistein1967uniform,duistermaat1974oscillatory}) for uniform asymptotic evaluation of 1D integrals with several nearby stationary points and singularities \cite{bleistein1967uniform}. The global mapping approach, while very elegant and systematic, leads to \emph{implicit} representations where the field is described in a generalized coordinate system that has to be calculated numerically. Alternatively, this mapping may be performed locally around any given point in configurational space, leading to \emph{explicit} field representations where the generalized coordinates are the local Cartesian coordinates. In this paper we  utilize the local mapping approach since our goal is to explore the wave mechanism of the 3D-AiB by deriving explicit field solutions where all the parameters are directly related to the aperture field distribution.  We start in Sec.~\ref{cat_type} by mapping the exact AiB spectral integral  in the vicinity of the curved beam axis, ending up with a  hyperbolic umbilic diffraction integral for the field. Next in Sec.~\ref{cat_finite_energy} we extend this integral to the case of \emph{finite-energy} AiB's by  adding a complex loss parameter into the phase of this integral.  In Sec.~\ref{cat_field} we calculate this integral by extending the closed-form series expansion of  \cite[Sec.~36.8]{DLMF} to  finite energy AiB's. Finally, in Sec.~\ref{num_example} we use this solution in order to check the validity of the conventional PE solution, and demonstrate that the latter suffers from a translational error due to an error in the beam axis under the PE approximation.

\section{Non Dispersive Airy Beams}  \label{non_dis_AiB}

In a previous paper \cite{KaganHeyman} we have dealt with the ray interpretation to the 2D-AiB and demonstrated that its curved propagation trajectory is a caustic of rays that emerge sideways from an extended aperture. Based on this concept we have extended the 2D-AiB to the ultra wide band (UWB) frequency domain and to the short-pulse time domain, by employing an initial field distribution that gives rise to a \emph{frequency-independent} ray skeleton, thus ensuring that all frequency components  propagate along the same trajectories and focus onto the same caustic \cite{KaganHeymanJOSA2011}.

\begin{figure}[tb]
\centering
      \subfigure[$z=0$]{%
      \label{Fig111}%
      \includegraphics[height=5cm]{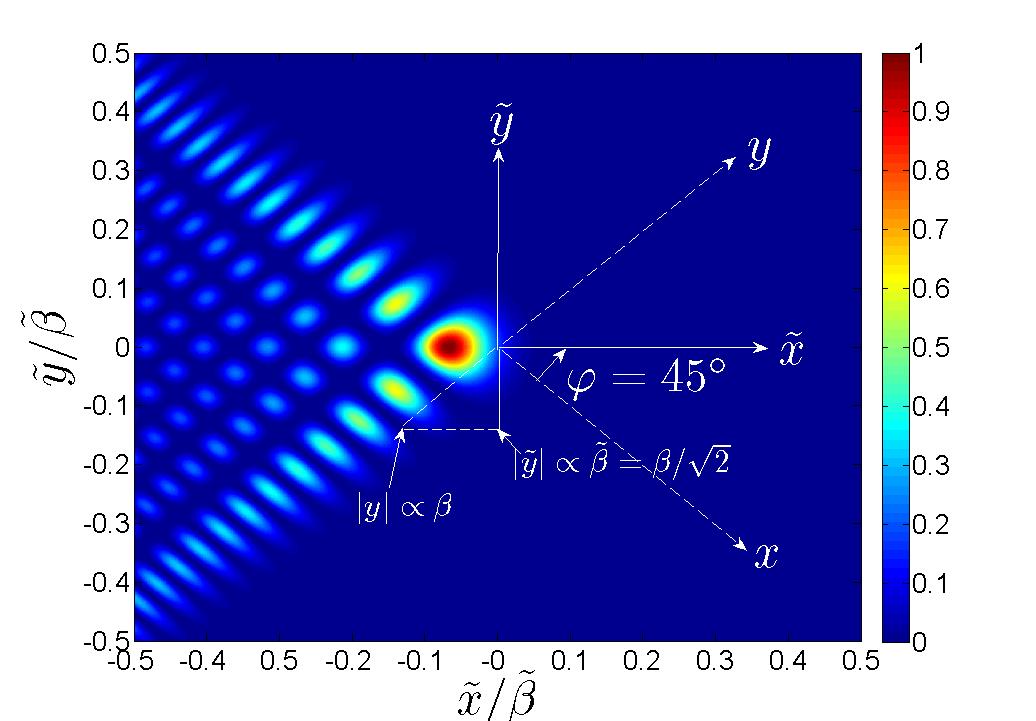}}%
       \subfigure[$z=0.8\tilde\beta$]{%
      \label{FigF222}%
      \includegraphics[height=5cm]{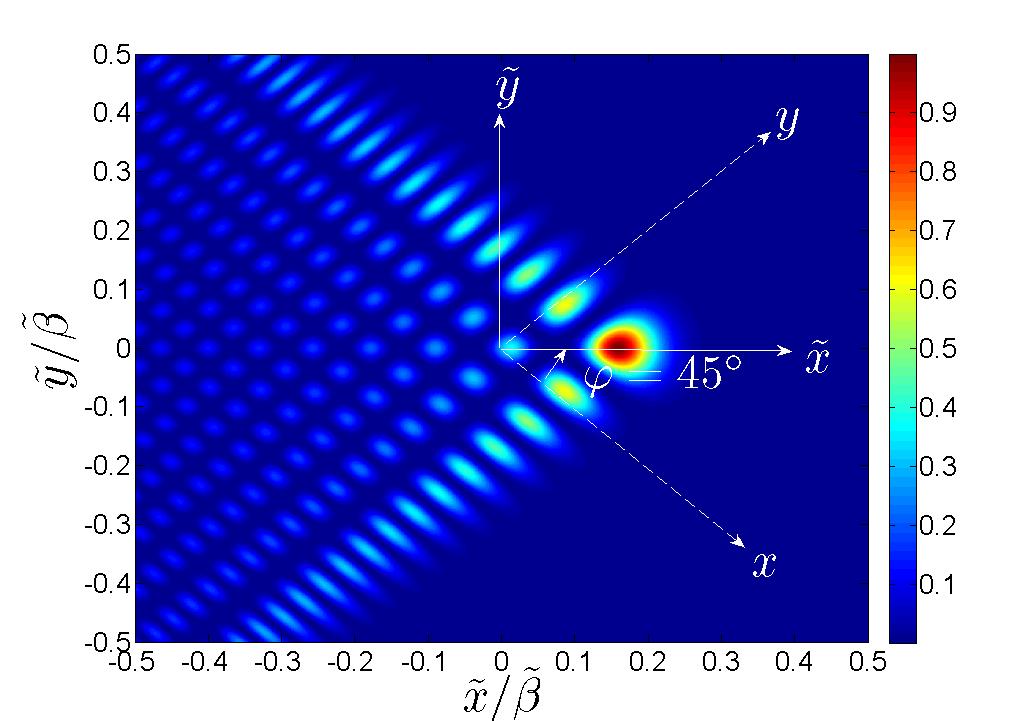}}%
\caption {(Color online)  Field intensity $|U|^2$ at (a) the aperture plane $z=0$, and (b) the $z=0.8\tilde\beta$ plane. The field is given by the PE solution in Eqs.~\eqref{gen_parax_field}--\eqref{parax_field}. Beam parameters are as in \eqref{beta_cond}--\eqref{alpha_cond} with $k\tilde\beta=100$, $\tilde\alpha=10^{-5}$. The beam propagates in the plane of symmetry $\tilde y=0$ in the direction $\varphi=45^\circ$. The two coordinate systems $(x,z)$ and $(\tilde x,\tilde y)$ are related via Eq.~\eqref{xy_tilde_sys}. The zeros $|U|=0$ in Fig.~(a) along the $x$ and $y$ axes are given, respectively, by $x=a_n\beta(\beta k)^{-2/3}$, $y=a_n\beta(\beta k)^{-2/3}$, where $a_n$ with $n=1,2,...$, are the zeros of Ai given in \cite[Table~9.9.1]{DLMF}.}
\label{Fig_uv_sys}
 \end{figure}

Extending this concept to the 3D case, the non-dispersive AiB field  in the half-space $z>0$ is generated by the aperture field distribution in the $z=0$ plane
\begin{align}
   U_{\scriptscriptstyle{0}}( x, y;\omega)= \exp[k(\alpha_{ x}\,  x+\alpha_{ y}\,  y)]\,\text{Ai}(\beta_{ x}^{-1/3}k^{2/3} x)\text{Ai}(\beta_{ y}^{-1/3}k^{2/3} y), \label{aperture_field}
\end{align}
where $( x, y)$ are the transversal coordinates, Ai is the Airy function, and a harmonic time-dependence $\exp(-i\omega
t)$ is assumed and suppressed throughout. $k=\omega/c$ is the wave-number and the wave-speed $c$ is assumed uniform. The exponential window in Eq.~\eqref{aperture_field} is added in order to render the energy of this distribution finite.  The parameters $\beta_{ x}, \beta_{ y}$, $\alpha_{ x}, \alpha_{ y}$ are \emph{frequency independent} scaling factors, with $\beta_{ x, y}$ controlling the widths of the field distribution along the $ x$ and $ y $ axes, respectively, and $\alpha_{ x, y}$  controlling the decay rates along these axes (see Fig.~\ref{Fig_uv_sys}).

It should be  noted that the conventional expression for the AiB initial condition of  Eq.~\eqref{aperture_field} is $U_\70(x,y)=\9{Ai}(x/x_\70)\9{Ai}(y/y_\70) \exp(a_x x+ a_y y)$\cite[Eq. (2)]{Broky08}. The scaling parameters in this expression are related to those in  Eq.~\eqref{aperture_field} via  $x_\70=\beta_{ x}^{1/3}k^{-2/3}$, $y_\70=\beta_{ y}^{1/3}k^{-2/3}$ and $a_{x,y}= k\alpha_{ x, y}$. We prefer the frequency independent $\beta_{ x, y}$ parameterization since it renders the ray skeleton and the propagation trajectory frequency independent (see Eq.~\eqref{beam_axis} and Sec.~\ref{caustic_top}).

The radiating field at $\3r=( x, y,z)$ due to the distribution in Eq.~\eqref{aperture_field} has a closed-form
solution within the parabolic wave equation (PE) which can be expressed as  \cite{Broky08}
\begin{align}
   &U( x, y,z)=F( x,z;\omega,\beta_{ x},\alpha_{ x}) F( y,z;\omega,\beta_{ y},\alpha_{ y})\exp(ikz),\label{gen_parax_field}
\end{align}
where $F$ is given by
\begin{align}
   &F(\rho,z;\omega,\beta, \alpha)= \text{Ai}\big[(k\beta)^{2/3}\big(\rho/\beta-(z/2\beta)^2+i\alpha z/\beta \big)\big]\nonumber \\
   & \qquad \qquad \qquad \quad \,\,\,\exp\big[ik(\rho z/2\beta -z^3/12\beta^2+\alpha^2z/2)\big] \nonumber \\
   &\qquad \qquad \qquad \quad \,\,\,\exp\big[k\alpha(\rho- z^2/2\beta)\big]. \label{parax_field}
\end{align}
The beam trajectory is defined by the vanishing of the arguments of the Airy functions in Eqs.~\eqref{gen_parax_field}--\eqref{parax_field}, and is given by\footnote{ Actually, the peak trajectory is described by replacing the $0$ on the right hand side of Eqs.~\eqref{beam_traj1}--\eqref{beam_traj2} by the asymptotically small parameter $(k\beta)^{-2/3}a'_1$ where $a'_1\simeq -1.0188$ is the first zero of $\text{Ai}'(x)$ \cite[Table~9.9.1]{DLMF}.}
\begin{align}
 & x/\beta_{ x}-(z/2\beta_{ x})^2=0 , \label{beam_traj1}  \\
 & y/\beta_{ y}-(z/2\beta_{ y})^2=0 . \label{beam_traj2}
\end{align}
Note that the $\beta_{x,y}$ parameterization implies that the beam trajectory is frequency independent.

For later use we introduce the coordinate system $(\tilde x,\tilde y,z)$  such that the beam trajectory  is contained in the $(\tilde x, z)$  plane (see Fig.~\ref{Fig_uv_sys}). This system  is obtained by rotating the $( x, y)$ system at the angle $\varphi=\tan^{-1}(\beta_{ x}/\beta_{ y})$
about the $z$-axis, viz
\begin{align}
\big(\tilde x,\tilde y\big)^\7T=
 \3R_{\varphi}
 \big( x, y\big)^\7T, \label{xy_tilde_sys}
\end{align}
where the superscript $T$ defines the transpose of a matrix and $\3R_{\varphi}$ is the rotation matrix
\begin{align}
\3R_{\varphi}=
\begin{pmatrix}  \cos\varphi  & \sin\varphi \\
                 -\sin\varphi & \cos\varphi
\end{pmatrix}. \label{rotation}
\end{align}

For simplicity, but without loss of generality, we shall consider the symmetrical case
\begin{align}
  &\beta_{ x} = \beta_{ y} \triangleq  \beta\triangleq\tilde \beta\sqrt{2}, \label{beta_cond} \\
  &\alpha_{ x}=\alpha_{ y} \triangleq \alpha \triangleq \tilde\alpha/\sqrt{2}, \label{alpha_cond}
\end{align}
such that $\varphi=45^\circ$ (see Fig.~\ref{Fig_uv_sys}). The parameters in the $(\tilde x,\tilde y)$ system are defined conveniently as $\tilde\beta$ and $\tilde\alpha$, and the beam trajectory (the ``beam axis'') is given by
\begin{align}
 \tilde x/{\tilde\beta}-\big({ z}/{2\tilde\beta}\big)^2=0.  \label{beam_axis}
\end{align}

The general case $\beta_{ x} \neq \beta_{ y}$ can be reduced to the symmetrical case  by the scaling  $ x \rightarrow  x [(\beta_{ x}^2+\beta_{ y}^2)^{1/2}/\beta_{ x}]^{1/3}$, $ y \rightarrow  y [(\beta_{ x}^2+\beta_{ y}^2)^{1/2}/\beta_{ y}]^{1/3}$ and applying the exponential windows in the new coordinate system.

The solution in Eqs.~\eqref{gen_parax_field}--\eqref{parax_field} is derived within the PE (paraxial) approximation and is therefore inaccurate at large distances where the AiB propagation trajectory curves. The ray solution of Secs.~\ref{field_calc}--\ref{catastrophe}, on the other hand, remains accurate even at  large distances since it is structured upon the ray skeleton of the \emph{exact} wave equation. The ray solution also provides a clear physical interpretation for the AiB and a tool for constructing other curving beam solutions in uniform and non-uniform media.

\section{Ray System and Caustic  Topology }\label{caustic_top}

\subsection{Ray System }\label{ray_sys1}

The properties of the AiB are fully described by the ray system that emerges from the aperture plane. We therefore start with the ray-system topology. Position $\3{r}=\big(x,y,z\big)^\7T$  along a ray can be expressed as
\begin{align}
  \3r=\3{r}'+\3{\hat s}\,\sigma , \label{ray_coord}
\end{align}
where $\3{r}'=\big(x' , y' , 0\big)^\7T$  is the ray initiation point in the $z=0$ plane, $\sigma$ is the distance along the ray, and  $\3{\hat{s}}$ defines the ray direction viz \begin{align}
 \3{\hat{s}}\triangleq\big(\xi,\,\eta,\,\zeta\big)^\7T\triangleq\big(\cos\theta_x,\,\cos\theta_y,\,\cos\theta_z\big)^\7T, \label{ray_direct}
\end{align}
with $\theta_{x,y,z}$ being the ray angles with respect to the $x$,$y$,$z$ axes, respectively, and
  \begin{equation}\label{zeta}
    \zeta=(1-\xi^2-\eta^2)^{1/2}.
\end{equation}

In order to find $\3{\hat{s}}$  for a given initiation point $(x',y')$ we substitute in Eq.~\eqref{aperture_field} the asymptotic form of the Airy function,
\begin{align}
\text{Ai}(x) \sim(-x\pi^2)^{-1/4}\sin\big[\frac{2}{3}(-x)^{3/2}+\frac{\pi}{4}\big], \quad x\ll -1, \label{Airy_asymp}
\end{align}
and recast the initial field in the form
\begin{align}
  U_\70(x',y') \approx \sum_{s=1}^4A_{\70s}\exp[ik\psi_{\70s}], \label{IC_for_rays}
\end{align}
where
\begin{align}
&\psi_{\70s}( x', y')=-2[\pm(- x')^{3/2} \pm(- y')^{3/2}]/3\beta^{1/2}, \label{psi_i} \\
& A_{\70s}( x', y')=\frac{i \beta^{1/6}}{4\pi k^{1/3}} \frac{e^{k\alpha(x'+y')}}{ (- x')^{1/4}(- y')^{1/4}}e^{-i\pi M_{\70s}/2} ,\label{A_0s}
\end{align}
with the $\pm$ signs in Eq.~\eqref{psi_i} taken as $++$, $+-$, $-+$, $--$, for $s=1,2,3,4$, respectively. In Eq.~\eqref{IC_for_rays}, $M_{\70s}=0$ for $s=1$, $M_{\70s}=1$ for $s=2,3$ and $M_{\70s}=2$ for $s=4$.
We assumed here that $\alpha$ is small so that the decay term  $\exp[k\alpha(x'+y')]$ is slowly varying on a wavelength scale and can be included in the amplitude $A_{\70s}$ rather than in the phase term $\psi_{\70s}$. This assumption is justified in Sec.~\ref{GO}.

\begin{figure}[tb]
\centering
      \subfigure[Ray species $s=1$ ]{%
      \label{Fig_S1}%
      \includegraphics[height=5.5cm]{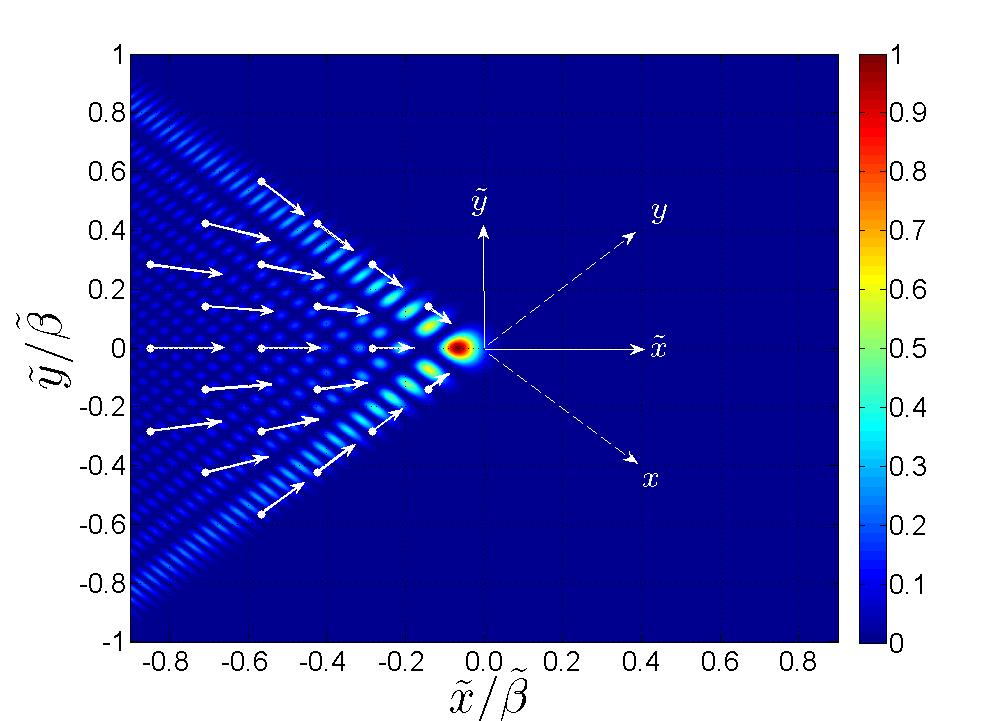}}%
       \subfigure[Ray species $s=2$]{%
      \label{Fig_S2}%
      \includegraphics[height=5.5cm]{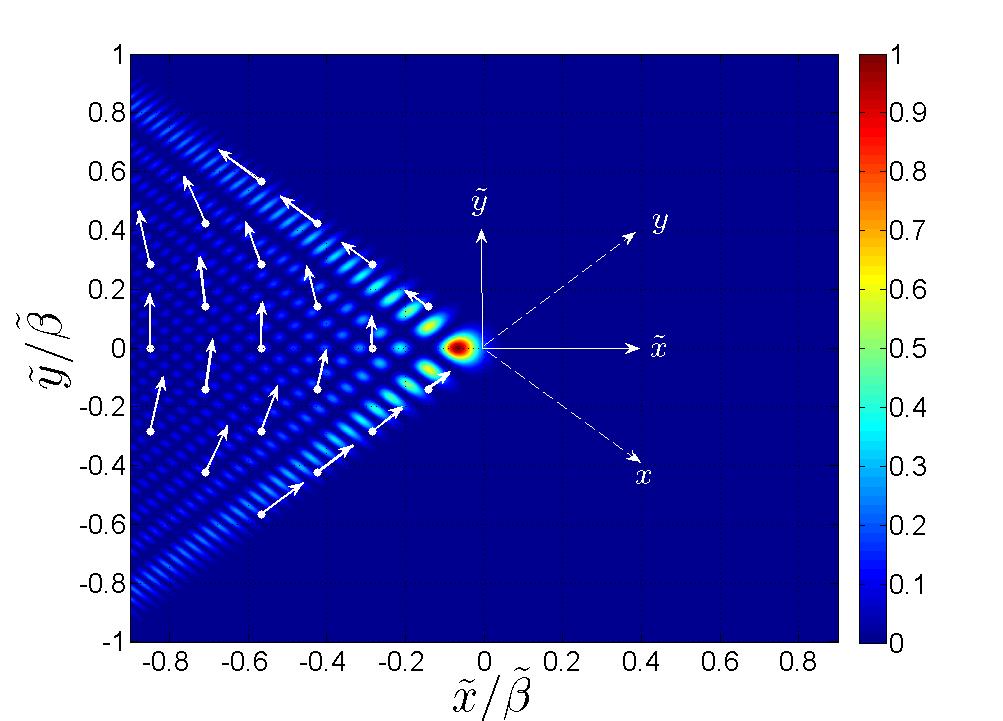}}%
      \\
       \subfigure[Ray species $s=3$]{%
      \label{Fig_S3}%
      \includegraphics[height=5.5cm]{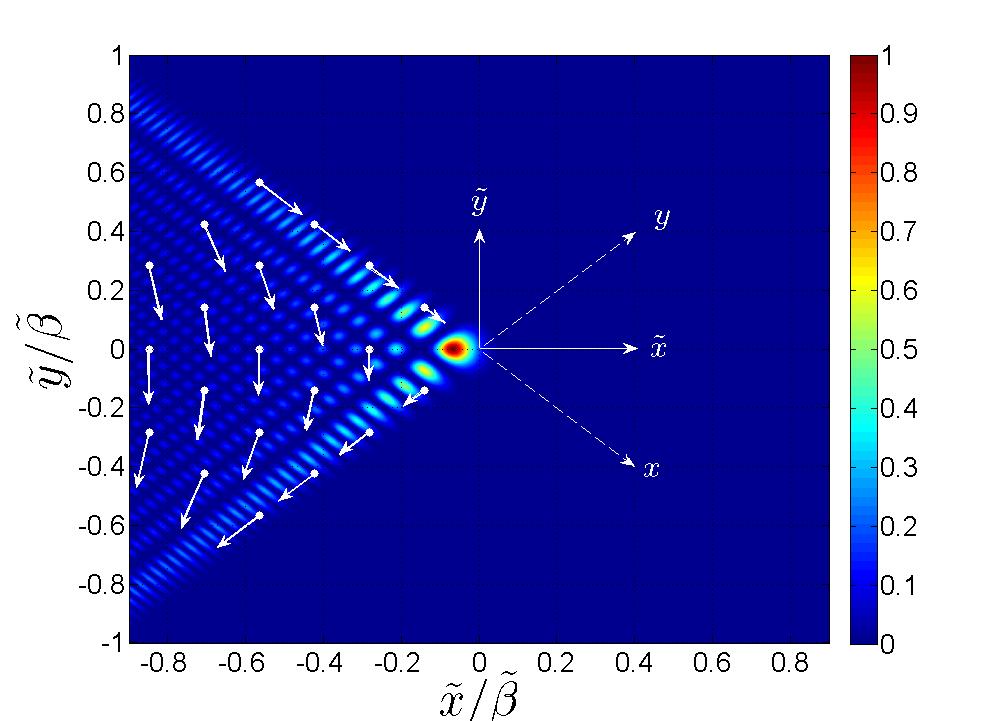}}%
       \subfigure[Ray species $s=4$]{%
      \label{Fig_S4}%
      \includegraphics[height=5.5cm]{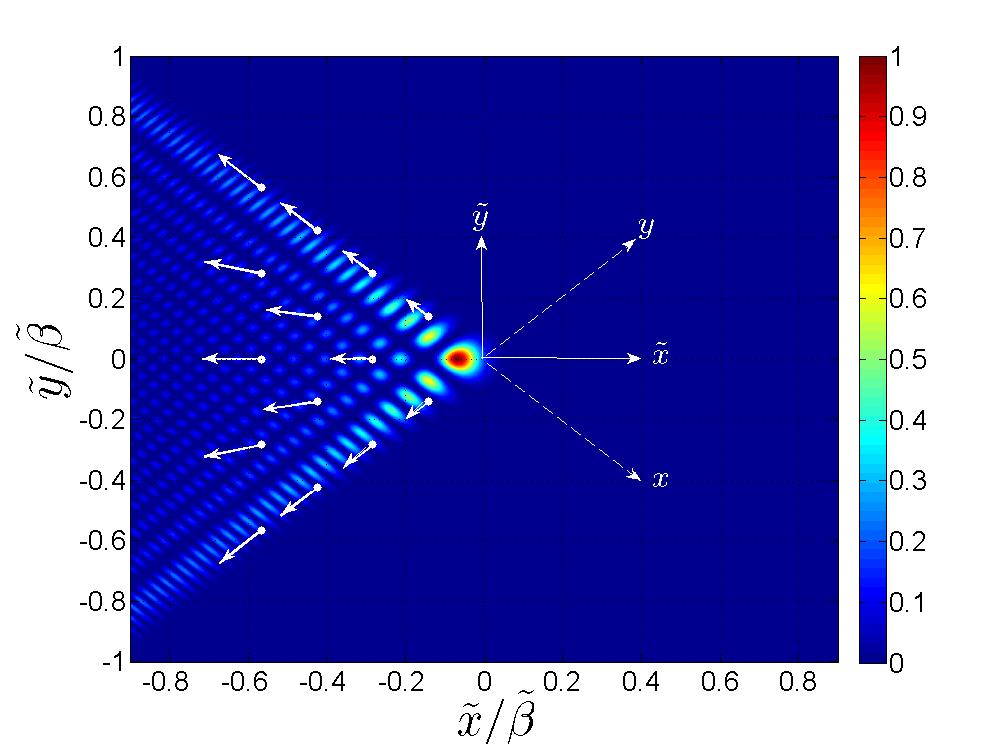}}%
\caption{(Color online)  Projections of the ray directions  $\3{\hat{s}}$ defined in Eqs.~\eqref{ray_direct} and \eqref{exit_angles} onto the $z=0$ plane (white arrows). Also shown for reference the intensity $|U|^2$ from Fig.~\ref{Fig111}. Figures (a)--(d) show the four  ray species defined in Eq.~\eqref{exit_angles} such that (a)  $\theta_x,\theta_y \in [0, \pi/2]$; (b)  $\theta_x\in [\pi/2, \pi]$ and $\theta_y \in [0,\pi/2]$ (c) $\theta_x \in [0,\pi/2]$ and $\theta_y\in [\pi/2, \pi]$  (d) $\theta_x,\theta_y\in [\pi/2,\pi]$.} \label{Fig_exit_angles}
 \end{figure}

One finds that each point $(x',y')$ in the aperture plane, with $ x', y'<0$, emits four rays, corresponding to $s=1,..,4$ in Eq.~\eqref{IC_for_rays}. The ray  directions $\3{\hat{s}}_s$ are obtained by local matching to the phase gradient viz
\begin{align}
 \big[\xi_s,\,\eta_s\big]=\big[\partial_{ x'},\,\partial_{ y'}\big]\psi_{\70s}(x',y')=\big[\pm(- x'/\beta)^{1/2},\,\pm(- y'/\beta)^{1/2}\big], \label{exit_angles}
\end{align}
where the ``$\pm$'' signs are taken as explained in Eq.~\eqref{psi_i}. Note that the ray directions are frequency independent if $ \beta$ is frequency independent. The directions of the four ray species projected onto the $z=0$ plane are depicted in Fig.~\ref{Fig_exit_angles}.
Finally, we note that the  result in Eq.~\eqref{exit_angles} has been derived asymptotically via Eq.~\eqref{Airy_asymp}, but as shown in Appendix~\ref{ray_eqs_appx}, it is also valid \emph{exactly}.

The ray trajectories in Eq.~\eqref{ray_coord} are parameterized in terms of the  ray coordinates $(\xi,\eta,\sigma)$ where the ray direction $(\xi,\eta)$ tags the initiation point $(x',y')$ via Eq.~\eqref{exit_angles}. The topology of the ray system is described by the Jacobian of the mapping from the $(x,y,z)$ coordinates to the ray coordinates $(\xi,\eta,\sigma)$, defined by
\begin{align}
 \5J= \frac{\partial(x,y,z)}{\partial(\sigma,\xi,\eta)}=
 \begin{Vmatrix}
 \3r_\sigma & \3r_{\xi} & \3r_{\eta}
\end{Vmatrix} ,\label{Jacobian}
\end{align}
where $\3r_\sigma=\big(\partial_\sigma x ,\, \partial_{\sigma} y,\, \partial_{\sigma} z\big)^\7T$ and $\| \, \|$ denotes the determinant. An explicit expression for $\5J$ is obtained by noting from Eq.~\eqref{exit_angles} that the exit point coordinates are defined by the ray coordinates via $(x',y')=-\beta(\xi^2,\eta^2)$. Substituting in  Eq.~\eqref{ray_coord} and calculating the derivatives in Eq.~\eqref{Jacobian} we obtain
\begin{align}
 \5J=
 \begin{Vmatrix}  \xi & \sigma-2\beta\xi & 0 \\
                \eta & 0 & \sigma-2\beta\eta   \\
                \zeta & -\sigma\xi/\zeta & -\sigma\eta/\zeta
\end{Vmatrix}. \label{J_app}
\end{align}
The  mapping is singular when $\5J=0$. This condition defines the \emph{envelope} of the ray system, also termed \emph{caustic} surface or \emph{catastrophe} \cite{nye1999natural,Kravtsov_catastrophe}. The structure of the caustic near the beam axis is discussed next in Sec.~\ref{caustic_top_parax}, with further details in  Appendix~\ref{exact_caustic_eqs}.  The complete caustic topology including also regions away from the beam axis is presented in  Sec.~\ref{caustic_complete}.

\subsection{Caustic Topology in the Vicinity of the Beam Axis} \label{caustic_top_parax}

\subsubsection{Caustic topology within the PE model} \label{caustic_top_parax_approx}

The PE approximation for the beam axis in Eq.~\eqref{beam_axis}  is valid only for moderate ranges before the axis curves too much. In this range one may approximate in Eq.~\eqref{zeta} $\zeta\simeq 1-\frac{1}{2}(\xi^2+\eta^2)$. Substituting into Eq.~\eqref{J_app} one obtains a closed form solution for the caustic equation $\5J=0$ (see Appendix~\ref{calc_parax_caustic}). Figure~\ref{Fig_caustic_parax} depicts the caustic structure within this approximation. It consists of two curved sheets joined at an edge that delineates the beam axis of  Eq.~\eqref{beam_axis}. Fig.~\ref{Fig_caustic_cross_z}  shows a cross sectional cut of the caustic in two constant-$z$ planes, where it is seen to form a right-angled corner. Figure~\ref{Fig_caustic_cross_z} also depicts the exact caustics to be discussed next.

\subsubsection{Exact caustic topology along the beam axis} \label{caustic_top_parax_exact}

Employing  \emph{exact} ray theory (see Appendix~\ref{calc_exact_caustic}) reveals further details of the caustic, which are also evident from the canonical catastrophe integral to be discussed in Sec.~\ref{catastrophe}. Here we discuss the structure of the caustic near the beam axis;  the complete \emph{global} structure beyond this zone will be discussed separately in Sec.~\ref{caustic_complete}.

The \emph{exact} caustic is shown in Fig.~\ref{Fig_caustic_exact}. It consists of two surfaces, referred to as ``caustic~1'' and ``caustic~2,'' each consisting of two sheets joined at an edge. Caustics~1 and~2  are very close  and hardly distinguishable on the figure scale. The caustic within the PE approximation is a degenerate case in which caustics~1 and~2 coalesce.
Figure~\ref{Fig_caustic_cross_z} zooms on cross-sectional cuts of the caustics at two constant-$z$ planes, showing that caustic~1 is a cuspoid, whereas caustic~2 is smooth, and that the distance between these caustics increases with $z$.  This topology is identified as a hyperbolic umbilic  diffraction catastrophe (see e.g., Fig.~4.4 in \cite{nye1999natural}) whose focal plane is located at the $z=0$ plane where the two surfaces coalesce along    the semi-infinite lines  $\{x=0, y<0\}$ and $\{y=0, x<0\}$ and form a right-angled corner at the origin.

The explosion of the PE caustic into a hyperbolic umbilic caustic at $z>0$ can also be expected from catastrophe stability theorems \cite{poston1996catastrophe,Kravtsov_catastrophe} as follows from the fact   that the right-angled corner at the cross-section of the PE caustic is  not a stable form (the only stable forms in 2D are the fold and the cusp; see e.g., Table 3.1 in \cite{nye1999natural}). Therefore, the corner will  not survive the perturbation  due to  exact propagation\footnote{ The PE model is a special case where  the effect of the propagation is only a translation (see Eqs.~\eqref{caustic_S1}-\eqref{caustic_S2}) and therefore the caustic  retains its unstable shape.}. Furthermore, the simplest stable singularity in 3D which contains a corner at a particular cross-section (termed focal plane) is the hyperbolic umbilic catastrophe (see also  \cite{berry1975cusped} where a similar argument has been used).
% 
% Using the exact formulation (see Appendix~\ref{calc_exact_caustic}) to calculate the \emph{global} structure of the caustic  reveals that further away from the aperture,   the hyperbolic umbilic caustic is no longer isolated and evolves into the more complex parabolic umbilic caustic (see further discussion in Sec.~\ref{caustic_complete}).}

Figure~\ref{Fig_caustic_cross_z} also shows that the deviation of the PE caustic from the exact one increases with $z$, thus limiting the range of validity for the  PE solution in Eqs.~\eqref{gen_parax_field}--\eqref{parax_field}.  In Fig.~\ref{Fig_edge_parax_vs_exact} we concentrate on the difference between the edges of the exact caustics~1 and~2 (Eqs.~\eqref{edge1_eq}--\eqref{vertex2_approx}) and of the PE caustic (Eq.~\eqref{beam_axis}) since this edge delineates the propagation trajectory of the AiB.

Figures~\ref{Fig_caustic1_ray_skeleton} and~\ref{Fig_caustic2_ray_skeleton} describe the ray formation of caustics~1 and~2, respectively, by depicting the rays emitted from points in the aperture plane and their point of tangency at the caustics. Actually, the figures show  traces of  the exit points in the aperture (thick lines), and the traces on the caustics of the corresponding points of tangency (thin lines with the arrows indicating the  direction of the traces). To emphasize the symmetry, traces with $\tilde y'\gtrless0$ appear as full or dashed lines, respectively. Lines of constant $\tilde y'$ and of constant $\tilde x'$ are plotted in magenta and blue, respectively.

As discussed in Eq.~\eqref{exit_angles} and Fig.~\ref{Fig_exit_angles}, each point in the aperture emits four rays, belonging to species $s=1,...4$. As seen in these figures, caustic~1 is formed by ray species~1,2,3, whereas caustic~2 is formed by species~1 only. Specifically, referring to Fig.~\ref{Fig_caustic1_ray_skeleton}, the rays of species~2 and~3 touch only caustic~1 (red points), whereas the rays of species~1 penetrate through caustic~1 (through the white circle), then touch caustic~2 (green point; see also Fig.~\ref{Fig_caustic2_ray_skeleton}), penetrate again through caustic~1 (white circle) and finally touch caustic~1 (red point).  In the limit $\tilde y'\to 0$, the rays of species~1 tend to the symmetry plane $\tilde y'=0$ thus giving rise to the somewhat counterintuitive situation in Fig.~\ref{Fig_caustic_coord} where the ray in the symmetry plane penetrates through the edge of caustic 1 and touches caustic~2, but then as it penetrates back through the cusped-edge of caustic~1 it is also tangent to the sheets that generate the cusp (see the blue solid-line trajectory in Fig.~\ref{Fig_caustic1_ray_skeleton}). This situation will be discussed again in connection with Figs.~\ref{Fig_rays} and~\ref{Fig_caustic_coord}.

\section{Exact Spectral Representation of the Field and Geometrical Optics} \label{field_calc}

\subsection{Spectral Representation} \label{spec}
Using the spectral representation of the Airy function \cite[Eq.~(9.5.1)]{DLMF}, the initial field in Eq.~\eqref{aperture_field} can be expressed as
\begin{align}
  &U_{\scriptscriptstyle{0}}( x, y;\omega)
  =\frac{\omega^{2/3}}{(2\pi)^{2}}\int_{-\infty}^{\infty} \int_{-\infty}^{\infty}d\xi d\eta A\, e^{i\omega \tau_\70(\xi,\eta)} e^{i\omega( \xi  x+\eta y)/c},  \label{init2}
\end{align}
where $\exp[i\omega(\xi  x+\eta y)/c]$ is the Fourier kernel, $(\xi,\eta)$ are the spectral variables associated with $( x, y)$ and
\begin{align}
  &\tau_{\scriptscriptstyle{0}}(\xi,\eta)=\beta[(\xi+i\alpha)^3+(\eta+i\alpha)^3]/3c  \label{tau_0} ,\\
  &A=(\beta/c)^{2/3}.  \label{A}
 \end{align}
The spectral integral in Eq.~\eqref{init2} has been expressed in a normalized \emph{non-dispersive} form \cite{STT1} where $\omega$ appears as a \emph{linear parameter} in the phase and $\tau{\scriptscriptstyle{0}}$ is \emph{frequency independent}. As a result $(\xi,\eta)$ have a pure \emph{frequency-independent} geometrical interpretation in terms of the plane wave angle (see discussion below), and $\tau{\scriptscriptstyle{0}}$ is the departure time of the $(\xi,\eta)$ plane wave from the $z=0$ plane. We use this normalization not only because its physical transparency but also since it may be extended to the time-domain \cite{KaganHeyman_3DAiB_part2_JOSA}.

The radiated field, obtained by adding the spectral propagator, is
\begin{align}
&U(\3r;\omega)=\frac{\omega^{2/3}}{(2\pi)^2} \int_{-\infty}^{\infty} \int_{-\infty}^{\infty}d\xi d\eta \, A\, e^{i\omega\tau(\xi,\eta)}   , \label{beam_freq} \\
&\tau(\xi,\eta;\3r)=\big[\tau_{\scriptscriptstyle{0}}(\xi,\eta)+\xi x+\eta y+ \zeta z\big]/c, \label{tau}
\end{align}
where $\zeta=\sqrt{1-\xi^2-\eta^2}$ is the spectral wave-number in the $z$-direction, chosen with $\text{Im}\,\zeta \geq 0$ for $\omega >0$. Consequently the
integration extends along the real $\xi,\eta$ axes from $-\infty$ to $\infty$ such that the $\xi$ contour  passes above and below the branch points $\xi=\mp\sqrt{1-\eta^2}$ corresponding to $\zeta$.

Equation~\eqref{beam_freq} expresses the field as a spectrum of plane-waves propagating in the direction $\3{\hat{s}}(\xi,\eta)$ defined in terms of $(\xi,\eta)$ via Eq.~\eqref{ray_direct}. Note that in  Sec.~\ref{caustic_top}, $(\xi,\eta)$ were parameters that define the ray in Eqs.~\eqref{ray_coord}--\eqref{ray_direct}, whereas in this section, they are  spectral variables that define the spectral-plane-wave direction. For simplicity we use the same notations and  distinguish between them from their context. Actually, the two are related since the ray coordinates of Sec.~\ref{caustic_top} are stationary points in the spectral $(\xi,\eta)$  domain, as explained after Eq.~\eqref{stationary_points} and also in Appendix~\ref{ray_eqs_appx}.

As noted after Eq.~\eqref{init2}, the spectral integral Eq.~\eqref{beam_freq} has a \emph{non-dispersive} form where $\omega$ appears as a \emph{linear parameter} in the phase and $\tau$ is \emph{frequency independent}, describing the arrival time of the plane wave that passes through $\3r$. This special form will be used in \cite{KaganHeyman_3DAiB_part2_JOSA} to evaluate the field of the Airy pulsed beam (AiPB).

The direct numerical evaluation of the spectral integral in Eq.~\eqref{beam_freq} is difficult since the integrand is very oscillatory at high frequencies. It may be evaluated asymptotically using the steepest descent method \cite{FM} which yields the geometrical optic (GO). Equation \eqref{beam_freq} will  be also used in Sec.~\ref{catastrophe} in order to calculate the AiB within the framework of Catastrophe theory.

\subsection{Geometrical Optics} \label{GO}
The main contributions in the integral of Eq.~\eqref{beam_freq} come from the vicinity of the saddle points, defined by
\begin{align}
 \partial_\xi\tau=0,\quad  \partial_\eta \tau=0, \label{stationary_points}
\end{align}
with $\tau$ given in Eq.~\eqref{tau}. We denote the solutions of Eq.~\eqref{stationary_points} as $(\xi_r,\eta_r)$ where $r$ is the \emph{ray-index}. For observation points far away from the caustic, the saddle points are distinct and the field can be expressed in terms of the isolated saddle point contributions \cite{FM}
 \begin{align}
 U\simeq\omega^{-1/3}\frac{A}{2\pi}\sum_r e^{-i(\pi/4)\mu_r} |H_r|^{-1/2}\,e^{i\omega\tau_r}, \label{sdp_field}
 \end{align}
with $\tau_r=\tau$ at $(\xi_r,\eta_r)$.  $H_r$ is the determinant of the Hessian matrix
\begin{align}
 \3H \triangleq
 \begin{pmatrix} \partial^2_\xi\tau   & \partial_\xi\partial_\eta\tau\\
                \partial_\eta\partial_\xi\tau  &\partial^2_\eta\tau
\end{pmatrix},  \label{Hessian}
\end{align}
evaluated at $(\xi_r,\eta_r)$, and $\mu_r$ is determined by the eigenvalues $h_{\7{1}r}$ and $h_{\7{2}r}$  of $\3H $, evaluated at $(\xi_r,\eta_r)$, via
 \begin{align}
 \mu_r=\9{sgn}\,h_{\71r}+\9{sgn}\,h_{\72r}. \label{eigen_H}
\end{align}
Expressions for $(\xi_r,\eta_r)$ and $\3H$ are given Eqs.~\eqref{d_xi_tau_gen}--\eqref{d_tau0_gen2} and~\eqref{tau_xi_xi}--\eqref{tau_xi_eta}, respectively. Expressions for $H_r$, $h_{\7{1,2}r}$ and $\mu_r$ are given in Eqs.~\eqref{H_eq3}--\eqref{eigen_H_app}.

As discussed in Eq.~\eqref{IC_for_rays}, $\alpha$ is relatively small and hence we consider first the case $\alpha=0$. In this case, the  saddle points $(\xi_r,\eta_r)$ corresponding to an observation points $\3r$ on the lit side of the caustic are real and equal to the ray parameters in Eq.~\eqref{ray_direct} of the rays reaching $\3r$ (see proof in Appendix~\ref{ray_eqs_appx}). The ray-index $r$ should  not be confused with the species-index $s$ introduced in Eq.~\eqref{IC_for_rays}: depending on $\3r$, the rays described  by the $r$th saddle points can be of any species.

The result in Eqs.~\eqref{sdp_field}--\eqref{eigen_H} can therefore be expressed in the GO form (see details in Appendix~\ref{go_field_appx})
\begin{align}
 &\tau_r=\tau_{\70r}+\sigma_r/c, \label{tau_i} \\
 &\omega^{-1/3}\frac{A}{2\pi}\frac{e^{-i(\pi/4) \mu_r}}{|H_r|^{1/2}}=e^{-iM_r\pi/2}A_{\70r}\sqrt{\bigg|\frac{\5J_r(0)}{\5J_r(\sigma_r)}\bigg|}, \label{H_i}
\end{align}
where $\sigma_r$ is  the ray length in Eq.~\eqref{ray_coord} and $\5J_r$ is the Jacobian in Eq.~\eqref{Jacobian}, both evaluated at $(\xi_r,\eta_r)$. All squared roots are taken as positive. $M_r$ is the Maslov index that counts the number of times ray $r$ has touched the caustic. $\tau_{\70r}$ and $A_{\70r}$ in Eqs.~\eqref{tau_i}--\eqref{H_i} are, respectively, the initial delay and amplitude of the $r$th ray, given by  Eq.~\eqref{psi_i}--\eqref{A_0s}, with $\tau_{\70r}=\psi_{\70s}/c$, both are taken  at $(x'_r,y'_r)$, the initiation point of the $r$th ray. Note that the GO amplitude in Eq.~\eqref{H_i} is invalid at the caustics where $\5J_r(\sigma_r)=0$.

Fig.~\ref{Fig_rays} depicts an illustrative example of the rays reaching a given observation point $\3r$. For simplicity we consider a point in the vicinity of  the beam trajectory where all rays are of species~1 (see Fig.~\ref{Fig_S1}). A point on the lit side of the caustic is reached by four rays, tagged by the index $r=1,2,3,4$.  The Maslov index of Eq.~\eqref{H_i} in this example is as follows: Ray $r=4$ has not touched the caustic  hence   $M_\74=0$; Rays $r=2,3$  have penetrated through ``caustic~1,'' then touched ``caustic~2'', and then penetrated again through ``caustic~1'' (see also ray~1 in Fig.~\ref{Fig_caustic1_ray_skeleton}), hence $M_\7{2,3}=1$; Ray $r=1$ is has penetrated through  ``caustic 1'', touched ``caustic 2'' and then touched the cusp at the edge of ``caustic 1''  while penetrating it (cf.~Fig.~\ref{Fig_caustic_coord}), so $M_\71=2$.  In addition to the four rays described above, the equation for the rays has three more solutions representing rays that emanate from the aperture at distant points where the field is very weak and the ray angles are very shallow with respect to the $z=0$ plane. These rays will not be considered therefore in the analysis.

Finally, we note that the GO field in Eq.~\eqref{sdp_field} has been derived via an asymptotic analysis of the exact spectral representation, but for $\sigma=0$ it reduces exactly to the asymptotic initial conditions in Eq.~\eqref{IC_for_rays} that were derived from the  asymptotic approximation of the Ai function in Eq.~\eqref{Airy_asymp}.

If $\alpha\neq0$ then the saddle points become complex. The saddle point analysis above still applies, but the physical interpretation is less transparent. However, since typically $\alpha$ is very small, one may expend $\tau_r$ and $H_r$ of Eq.~\eqref{sdp_field} into Taylor series about the real saddle points $(\xi_r,\eta_r)$ for $\alpha=0$. Keeping terms to leading order in $\alpha$  one  obtains the result in Eq.~\eqref{sdp_field} with Eqs.~\eqref{tau_i}--\eqref{H_i} except that here $A_{\70r}\to A_{\70r} e^{k\tilde\alpha \tilde{x}'_r}$ (cf.~Eq.~\eqref{A_0s}; see details in Appendix~\ref{go_field_appx}).

 \subsection{The field Near the Caustic }  \label{ugo}
 Near the caustic, the ray-system Jacobian vanishes and the GO amplitude in Eq.~\eqref{H_i} explodes. From a spectral perspective,  the relevant saddle points come close together and the second order asymptotic analysis in Eqs.~\eqref{sdp_field}--\eqref{eigen_H} is no longer valid. One should resort, instead, to  higher order asymptotics that may accommodate two or more nearby saddle points.  The simplest situation occurs near the smooth parts of the caustics which are fold catastrophe where two saddle points coalesce. In these regions, the field may be described uniformly via the uniform geometrical optics (UGO) \cite{Ludwig65,Kravtsov64} where the field is expressed explicitly in terms of the ray-parameters (i.e., the phase and the Jacobian), but these parameters are combined together so that the solution remains regular near and on the caustic while reducing to the conventional GO expressions further away from it. The simple UGO is not valid near the beam trajectory where the caustic topology is a hyperbolic umbilic as discussed in Sec.~\ref{caustic_top_parax_exact} and Figs.~\ref{Fig_caustic_cross_z}--\ref{Fig_caustic2_ray_skeleton}.  The field in that region is considered next in Sec.~\ref{catastrophe}.

\section{The Field Near the Propagation Trajectory} \label{catastrophe}

Near the beam trajectory, the caustic topology is that of a hyperbolic umbilic, hence, as explained in the introduction, we refer to the general framework of catastrophe theory in order to calculate the field. In this approach, the field is represented by the canonical catastrophe integral that can be written  as
\begin{align}
&\bar U(\bar x,\bar y,\bar z)=\frac{C_\70}{(2\pi)^2} \int\hspace{-0.5em}\int d \bar\xi d\bar\eta\,  e^{i\Phi(\bar\xi,\bar\eta;\bar x,\bar y,\bar z)}, \label{cat_spec}
\end{align}
where $( \bar x,\bar y,\bar z)$ are generalized coordinates in the vicinity of the caustic such that $\bar z$ is a coordinate along the caustic and $( \bar x,\bar y)$ are transversal coordinates, and $(\bar \xi,\bar \eta)$ are the spectral variables associated with $( \bar x,\bar y)$. $\Phi$  has a canonical polynomial form for each elementary catastrophe (see e.g., Eq.~\eqref{phi_canon} for the hyperbolic umbilic). Near the caustic, the integral in Eq.~\eqref{cat_spec} can be evaluated by standard numerical techniques such as series expansion \cite[Sec.~36.8]{DLMF} (see also discussion in Sec.~\ref{cat_field}), whereas further away from the caustic, it reduces asymptotically via saddle point analysis to the GO field. The spatial and spectral coordinates in Eq.~\eqref{cat_spec} are typically normalized to be dimensionless and $C_\70$ is a normalization constant such that $\bar U=1$ at $(\bar x,\bar y, \bar z)=0$.

In the theory of diffraction catastrophes \cite{berry1980iv,berry1976waves}, the exact integral representation of the field, expressed either spatially (i.e., via a Kirchhoff-type integral) or spectrally, is mapped asymptotically via a method outlined in \cite[Sec.~5]{berry1976waves} into the canonical integral in Eq.~\eqref{cat_spec}, thus leading to a \emph{globally uniform} solution of the field.\footnote{Specifically, in the present case, one applies a change of variables such that the  stationary points of the phase function $\tau$ in the exact spectral integral in Eq.~\eqref{beam_freq} are mapped to those of the phase $\Phi$ in Eq.~\eqref{cat_spec}, while satisfying the constraint that the phases at the respective points are equal, i.e., $\omega\tau=\Phi$.}

As discussed in the introduction, in the \emph{globally uniform} approach, the generalized coordinates in $\Phi$ are related \emph{implicitly} to the geometrical coordinates near the caustic. Our goal, however, is to obtain an \emph{explicit} mapping such that these generalized coordinates are directly related to the local Cartesian coordinates along the beam trajectory as shown in Fig.~\ref{Fig_caustic_coord}, and also to the  parameters of the aperture field distribution. Such explicit expressions are obtained by the \emph{locally uniform} approach of Sec.~\ref{cat_type} where we concentrate on the region near the caustic. In this approach, we transform the exact spectral integral of Eq.~\eqref{beam_freq} to the local Cartesian coordinate system in Fig.~\ref{Fig_caustic_coord}, and then approximate the spectrum paraxially around the caustic direction. This leads directly to the spectral integral in Eq.~\eqref{cat_spec} where the phase is recognized as hyperbolic umbilic. This integral describes the field in the transition layer around the caustic, but its validity range is much larger and in particular it reduces to the GO field away from the caustic. As discussed in the introduction, this direct approach also provides a convenient tool for the synthesis of other types of beams.

As was done in Sec.~\ref{caustic_top}, we consider first in Sec.~\ref{cat_type} the case where $\alpha=0$ where the rays are real. The modifications for the finite energy AiB where $\alpha\neq0$ are done in Sec.~\ref{cat_finite_energy}. In Secs.~\ref{cat_field}--\ref{num_example} we discuss the numerical evaluation of the integral.

\subsection{Mapping the Exact Spectral Integral to the Canonical Form} \label{cat_type}

The canonical integral is constructed about a a typical reference point along the beam trajectory, which is taken to be on the edge of caustic~2 (see  Fig.~\ref{Fig_caustic_coord}; recall from the discussion in Sec.~\ref{caustic_top} that the caustic topology near the beam axis is hyperbolic umbilic, consisting of two sheets denoted as caustic~1 and~2, each has an edge in the symmetry plane $\tilde y=0$, see also Figs.~\ref{Fig_caustic_cross_z}--\ref{Fig_caustic_exact}). This reference point is parameterized, conveniently, by $\theta_z$, the angle of the tangent to the edge at this point with respect to the $z$-axis. The coordinates at this point are given by $(\tilde x_\72(\theta_z),0, z_\72(\theta_z))$ of  Eq.~\eqref{edge2_coord} (see also Eqs.~\eqref{edge2_eq} and~\eqref{vertex2_approx}). The Cartesian coordinates around this point, denoted as $(\hat x,\hat y,\hat z)$, are chosen such that $\hat z$ is the tangent to the edge and  $(\hat x,\hat y)$ are the normal and the bi-normal, respectively (see Fig.~\ref{Fig_caustic_coord}), and are given, therefore, by
\begin{align}
 \big(\hat x,\,\hat z\big)^\7T=
 \3R_{\theta_z}
 \big(\tilde x-\tilde x_\72,\, z- z_\72\big)^\7T, \qquad \hat y=\tilde y,  \label{caustic_coord_sys}
\end{align}
where $\3R_{\theta}$ is the rotation matrix in Eq.~\eqref{rotation} with $\varphi=\theta_z$.

The spectral integral in the $(\hat x,\hat y,\hat z)$ system is obtained from the one in Eq.~\eqref{beam_freq}. We start by expressing the latter in the   $(\tilde x,\tilde y)$ system of Eq.~\eqref{xy_tilde_sys} by rotating the spectral variables such that $\big(\tilde\xi,\,\tilde\eta\big)^\7T=\3R_{45^0}\big(\xi,\,\eta\big)^\7T$ with $\3R$ given in Eq.~\eqref{rotation}. Substituting into Eqs.~\eqref{beam_freq}--\eqref{tau} we obtain
\begin{align}
U(\tilde{\3r};\omega)=\frac{\omega^{2/3}}{(2\pi)^2} \int\hspace{-0.5em}\int d\tilde\xi d\tilde\eta\, \tilde A\, e^{i\omega\tau(\tilde\xi,\tilde\eta)}   , \label{beam_freq_tilde}
\end{align}
where
\begin{align}
  &\tau(\tilde\xi,\tilde\eta)=\tilde\beta[(\tilde\xi+i\tilde\alpha)^3/3+(\tilde\xi+i\tilde\alpha)\tilde\eta^2]/c+\tilde\xi \tilde x+\tilde\eta\tilde y+ \tilde\zeta z,\label{tau_2}\\
  &\tilde A=(\tilde\beta/2c)^{2/3}, \label{A2}
 \end{align}
with $\tilde\zeta=(1-\tilde\xi^2-\tilde\eta^2)^{1/2}$. Recall that here we assume $\tilde\alpha=0$.

The spectral variables $(\hat\xi,\hat\eta,\hat\zeta)$ corresponding to $(\hat x,\hat y,\hat z)$ are related to $(\tilde\xi,\tilde\eta,\tilde\zeta)$ via
\begin{align}
 \big(\hat\xi,\,\hat\zeta\big)^\7T =
 \3R_{\theta_z}
 \big(\tilde\xi,\,\tilde\zeta\big)^\7T, \quad \hat\eta=\tilde\eta,   \label{rotate_spec}
\end{align}
with $\hat\zeta=(1-\hat\xi^2-\hat\eta^2)^{1/2}$. Substituting Eqs.~\eqref{caustic_coord_sys} and~\eqref{rotate_spec} into \eqref{tau_2}, expanding $\hat\zeta$ to a Taylor series about $(\hat\xi,\hat\eta)=(0,0)$, and keeping terms up to third order, we obtain the following expression for $\tau$
\begin{align}
 \tau=\tau_a+\tilde\beta D_\71\hat\xi^3/c+\tilde\beta D_\72\hat\xi\hat\eta^2/c+\hat\xi \hat{x}/c+\hat\eta \hat y/c+\hat z/c-\hat\xi^2\hat z/2c-\hat\eta^2(\hat z-\hat\delta)/2c, \label{tau_rotate_coord}
\end{align}
where
\begin{align}
 &\tau_a=2\tilde\beta\sin\theta_z(1-4\sin^2\theta_z/3)/c, && \hat\delta=2\tilde\beta\sin^3\theta_z, \label{tau_v} \\
 &D_\71=\cos\theta_z(5\cos^2\theta_z-3)/6, && D_\72=\cos\theta_z(1+\cos^2\theta_z)/2. \label{D1}
\end{align}
 $\tau_a$ in Eq.~\eqref{tau_v} is the spectral delay at $\hat z=0$ for $(\hat\xi,\hat\eta)=\30$. Referring to Fig.~\ref{Fig_caustic_coord}, $\hat\delta$ is identified as the distance along the tangent from the reference point $(\hat x,\hat y,\hat z)=\30$ to the cusped edge of caustic~1. From Eq.~\eqref{rotate_spec} $d\tilde\xi$ in Eq.~\eqref{beam_freq_tilde} is expressed as $d\tilde\xi=(\cos\theta_z  -\sin\theta_z\,\hat\xi/\hat\zeta)\, d\hat\xi\approx \cos\theta_z\,d\hat\xi$ where the last term is the approximation for $\hat\xi=0$. For $\theta_z\ll1$ one may use $D_\71\approx \frac{1}{3}-\theta_z^2$
and $D_\72\approx 1-\theta_z^2$.

Concentrating only on the $\hat z=0$ plane perpendicular to the edge, and normalizing the coordinates, the integral in Eq.~\eqref{beam_freq_tilde} with the phase $\tau$ of Eq.~\eqref{tau_rotate_coord} can be written in the form
 \begin{align}
 U=\bar A e^{i\omega\tau_a}\,\bar U(\bar x,\bar y;  \bar \delta)\cos\theta_z, \label{field_cat}
\end{align}
where $\tau_a$ is given in  Eq.~\eqref{tau_v} and  $\bar U$ is the canonical integral in Eq.~\eqref{cat_spec} with
\begin{align}
&\Phi (\bar\xi,\bar\eta;\bar x,\bar y;\bar \delta)=\bar\xi^3+\bar\xi\bar\eta^2+\bar\xi \bar x+\bar\eta \bar y +\bar\eta^2 \bar\delta,\label{phi_canon}\\
& C_\70=3^{1/6}2^{4/3}/[\text{Ai}(0)]^2.
\end{align}
The normalized spatial and spectral coordinates in Eq.~\eqref{phi_canon} are
\begin{align}
&\bar x=(k\tilde\beta)^{2/3}D_\71^{-1/3}\, \hat x/\tilde\beta , &&\bar y=(k\tilde\beta)^{2/3}D_\71^{1/6}D_\72^{-1/2}\,\hat y/\tilde\beta, \label{caustic_coord}\\
&\bar\xi= (k\tilde\beta)^{1/3}D_\71^{1/3}\,\hat\xi, &&\bar\eta=(k\tilde\beta)^{1/3}D_\71^{-1/6}D_\72^{1/2}\,\hat\eta, \label{caustic_spec}
\end{align}
and the normalized parameters are
\begin{align}
&\bar\delta=(k\tilde\beta)^{1/3}D_\71^{1/3}D_\72^{-1}\hat\delta/2\tilde\beta=(k\tilde\beta)^{1/3}D_\71^{1/3}D_\72^{-1}\sin^3\theta_z \approx (k\tilde\beta/3)^{1/3}\theta^3_z , \label{caustic_param}\\
&\bar A=D_\71^{-1/6}D_\72^{-1/2}/C_\70,\label{caustic_param-A}
\end{align}
where in the second equality in $\bar\delta$ we substituted $\hat \delta$ from Eq.~\eqref{tau_v}, while the last term applies for $\theta_z\ll1$.

The spectral phase $\Phi$ in Eq.~\eqref{phi_canon}, which is the \emph{simplest} polynomial that describes the AiB  in the vicinity of the caustic, is of the hyperbolic umbilic type \cite[pp. 66]{nye1999natural},\cite[pp.~41]{Kravtsov_catastrophe}. It follows that the field-scale near the caustic is governed by the factor $(k\tilde\beta)^{2/3}$ that scales  the local coordinates $(\hat x/\tilde\beta,\hat y/\tilde\beta)$ in Eq.~\eqref{caustic_coord}. Note that the $k^{2/3}$ scaling is expected in view of the general theory of diffraction catastrophes \cite[Sec.~36.6]{DLMF}.

The caustic topology is controlled by the parameter  $\bar\delta$ of Eq.~\eqref{caustic_param} which represents the normalized distance between caustics~1 and~2 such that it scales with frequency like $(k\tilde\beta)^{1/3}$ and also grows with the observation range $z$. For  $\bar\delta \gg 1$,  caustics~1 and~2 are far apart and may be described by separated cusp and fold  catastrophes, respectively. In this case, the hyperbolic umbilic integral may be reduced to the Pearcey and Airy integrals \cite[Sec.~36.2(ii)]{DLMF} that represent, respectively, the field near these catastrophes.

Equation~\eqref{field_cat} describes the AiB field in a plane normal to the beam axis. The solution in the entire domain along the axis is obtained by tracking  the local $(\hat x,\hat y, \hat z)$ coordinate system of Eq.~\eqref{caustic_coord_sys} along the beam axis.  The parameters in Eqs.~\eqref{tau_v}--\eqref{D1}  are  therefore expressed in terms of $\theta_z(z)$ which describes points on the beam trajectory (see Eq.~\eqref{edge2_coord}). Since this local spectrum representation is constructed entirely using the ray system topology near the caustic, it follows that the ray approach can be used to construct AiB-type solutions that follow any prescribed convex trajectory by back-projecting the rays to construct the relevant aperture distribution (e.g., the 2D example in \cite{AiB_arbitrary_traj_PRL2011}).

There are  several alternative forms for the hyperbolic umbilic spectral phase $\Phi$, all sharing the ``germ'', i.e., the highest order terms of $\Phi$ which depend only on $(\bar\xi,\bar\eta)$ and define the catastrophe. The phase $\Phi$ in Eq.~\eqref{phi_canon} contains the germ   $\bar\xi^3+\bar\xi\bar\eta^2$, but there is another form used in \cite[Sec.~36.3]{DLMF} which is discussed in Appendix~\ref{Alt_canon}. The rest of the phase terms, called   perturbation terms (or  unfolding terms), can have different forms \cite[chap.~4]{gilmore1993catastrophe} without changing the catastrophe type, as long as they are small. For example, the phase used in \cite[pp.~41]{Kravtsov_catastrophe} is the same as in Eq.~\eqref{phi_canon}  with  $\bar\eta^2\bar\delta$ replaced by $\bar \xi^2\bar\delta$, but since these are perturbation terms, the two forms are equivalent. In fact, the latter can be obtained by setting the reference point of the local coordinates on the edge of caustic~1 and taking the $\hat z$ axis along the direction of the ray that touches the cusped edge there (see point $\hat z=\hat\delta$ in  Fig.~\ref{Fig_caustic_coord}). We prefer the form in Eq.~\eqref{phi_canon} since in the alternative one, $\hat z$ is not tangent to the edge of caustic 1 (see dashed line in Fig.~\ref{Fig_caustic_coord}) and the tracking along this edge becomes more complicated. Finally, in Sec.~\ref{cat_finite_energy} below we extend $\Phi$ to the finite energy AiB case.

\subsection{Finite Energy AiB} \label{cat_finite_energy}

Referring to Eq.~\eqref{tau_2}, we now consider the case where  $\alpha\neq0$. Repeating the procedure thereafter we obtain a polynomial approximation for $\tau$ which is similar to  Eq.~\eqref{tau_rotate_coord}
\begin{align}
 \tau=\tau_a^\alpha+ \tilde\beta D_{\71}^\alpha\,\hat\xi^3/c+ \tilde\beta D_{\72}^\alpha\,\hat\xi\hat\eta^2/c+\hat\xi\hat{x}^\alpha/c+\hat\eta \hat y/c+\hat z/c-\hat\xi^2(\hat z-\epsilon)/2c-\hat\eta^2(\hat z-\hat\delta^\alpha)/2c, \label{tau_alpha}
\end{align}
except that the coordinate $\hat x$ of Eq.~\eqref{caustic_coord_sys} and the coefficients $\tau_a$, $D_\7{1,2}$ and $\hat \delta$ of Eqs.~\eqref{tau_v}--\eqref{D1} are now replaced by
\begin{align}
& \hat x^\alpha =\hat x+\tilde\beta[i\tilde\alpha\sin(2\theta_z)-\tilde\alpha^2\cos\theta_z],\label{x_alpha}\\
&\tau_a^\alpha=\tau_a+\tilde\beta[i\tilde\alpha\sin^2\theta_z-\tilde\alpha^2\sin\theta_z-i\tilde\alpha^3/3]/c, \label{tau_v_alpha}\\
&\hat\delta^\alpha=\hat\delta+\tilde\beta[ i\tilde\alpha+i\tilde\alpha\cos^2\theta_z+\tilde\alpha^2\sin\theta_z], \label{delta_alpha}\\
& D_\7{1,2}^\alpha = D_\7{1,2}-i\tilde\alpha\sin\theta_z\cos\theta_z/c, \label{D1_alpha}
\end{align}
and the term  $\hat\xi^2\hat z$ in Eq.~\eqref{tau_rotate_coord} now becomes $\hat\xi^2(\hat z-\epsilon)$ with
\begin{align}
 \epsilon=\tilde\beta(i\tilde\alpha+\tilde\alpha^2\sin\theta_z/2-i\tilde\alpha\sin^2\theta_z/2).
\end{align}
For $\hat z=0$, $\Phi$ of Eq.~\eqref{phi_canon} becomes
\begin{align}
 \Phi(\bar\xi,\bar\eta;\bar x,\bar y;\bar \delta;\bar\epsilon)=\bar\xi^3+\bar\xi\bar\eta^2+\bar\xi \bar x+\bar\eta \bar y +\bar\xi^2\bar\epsilon+\bar\eta^2\bar \delta , \label{phi_canon_finite}
\end{align}
where
\begin{align}
\bar\epsilon=(k\tilde\beta)^{1/3}(D_\71^\alpha)^{-2/3}\epsilon/\tilde\beta,
\end{align}
and the other normalized variables are given by Eqs.~\eqref{caustic_spec}--\eqref{caustic_param} with $\hat x$,  $\hat\delta$, $\tau_a$, $D_\7{1,2}$ replaced by the corresponding parameters $\hat x^\alpha$, $\hat\delta^\alpha$, $\tau_a^\alpha$, $D^\alpha_\7{1,2}$   of Eqs.~\eqref{x_alpha}--\eqref{D1_alpha}. Note that Eq.~\eqref{phi_canon_finite} reduces back to Eq.~\eqref{phi_canon} for $\tilde \alpha=0$.

Expression~\eqref{field_cat} for the field becomes
\begin{align}
 U=\bar A\, e^{i\omega\tau_a^\alpha}\,\bar U(\bar x,\bar y;  \bar \delta;\bar\epsilon)\cos\theta_z, \label{field_cat_finite}
\end{align}
where  $\bar U$ is the canonical integral in Eq.~\eqref{cat_spec} with $\Phi(\bar\xi,\bar\eta;\bar x,\bar y;\bar \delta;\bar\epsilon)$ given in  Eq.~\eqref{phi_canon_finite}.

The expression in Eq.~\eqref{phi_canon_finite} has  an additional perturbation term $\bar\xi^2\bar\epsilon$ (assuming $\tilde\alpha$ is small ), but as noted at the end of Sec.~\ref{cat_type},   the diffraction pattern of the field will remain hyperbolic umbilic, as can be verified both theoretically and experimentally \cite{Sivil_Phyrev,Broky08}. The main effect of $\tilde\alpha$ is an overall amplitude decay which is mainly due to the term $e^{i\omega\tau_a^\alpha}$ in Eq.~\eqref{field_cat_finite}.

%%%%%%%%%%%%%%%%%%%%%%%%%%%%%%%%

%%%%%%%%%%%%%%%%%%%%%%%%%%%%%%

\subsection{Numerical Evaluation of the Canonical Integral} \label{cat_field} 

 The canonical integral in Eq.~\eqref{cat_spec} with $\Phi$ in Eq.~\eqref{phi_canon} or~\eqref{phi_canon_finite} can be evaluated by a straightforward numerical integration. To accelerate the convergence, the integration contours for $|\xi|,|\eta|\gg1$ can be deformed into the complex plane  so as to reach infinity along the asymptotic valleys of $\exp(i\Phi)$. Fastest convergence is obtained if the $\xi$-integration starts asymptotically along the line $\arg \xi=5\pi/6$ and ends along the line $\arg \xi=\pi/6$. Likewise, the $\eta$ integration may start asymptotically along the line $\arg \eta=-3\pi/4$ and ends along the line $\arg \eta=\pi/4$.  Another approach is to use the 1D integral representation for the hyperbolic umbilic as explained in \cite[Sec.~36.15(iii)]{DLMF}.

Here we shall use the standard series expansion for the canonical double  integral in \cite[Eq.~36.8.3]{DLMF} and extend it to the case of finite energy AiB. The small parameters in the series expansion are  $\bar \delta   \sim O((k\tilde\beta)^{1/3}\sin^3\theta_z)$ and $\bar \epsilon \sim O((k\tilde\beta)^{1/3}\tilde\alpha)$. For typical values of $\tilde\alpha, \omega, \tilde\beta$ such that $\bar \delta,\bar\epsilon \ll 1$ we may expand into a Taylor series the term $\exp[ i(\bar\xi^2\bar\epsilon+\bar\eta^2\bar\delta)]$ appearing in the expression for $\bar U$ in Eq.~\eqref{cat_spec} with $\Phi$ given in  Eq.~\eqref{phi_canon_finite}. We obtain
\begin{align}
&\bar U(\bar x,\bar y; \bar \delta;\bar \epsilon )=\frac{C_\70}{(2\pi)^2}\sum_{p,m}(-i)^{(p+m)}(p!m!)^{-1}\bar \epsilon^{p}\,\bar \delta^{m} \int\hspace{-0.5em}\int d \bar\xi d\bar\eta\, (i\bar\xi)^{2p}\,(i\bar\eta)^{2m} e^{i\bar\Phi(\bar\xi,\bar\eta;\bar x,\bar y)}  , \label{cat_spec2}
\end{align}
where
\begin{align}
 \bar\Phi(\bar\xi,\bar\eta;\bar x,\bar y)&=\Phi(\bar\xi,\bar\eta;\bar x,\bar y;\bar \delta;\bar \epsilon)-\bar\xi^2\bar\epsilon-\bar\eta^2\bar \delta\nonumber\\
 &=\bar\xi^3+\bar\xi\bar\eta^2+\bar\xi \bar x+\bar\eta \bar y. \label{phi}
\end{align}
Using $(i\bar\xi) \leftrightarrow \partial_{\bar x}$,$(i\bar\eta) \leftrightarrow \partial_{\bar y}$  in Eq.~\eqref{cat_spec2} and taking the derivatives outside of the integral, we rewrite Eq.~\eqref{cat_spec2} as
\begin{align}
&\bar U(\bar x,\bar y,\bar z)=C_\70\sum_{p,m}(-i)^{(p+m)}(p!m!)^{-1}\bar \epsilon^{p}\,\bar \delta^{m}\,\partial^{2p}_{\bar x}\partial^{2m}_{\bar y} I(\bar x,\bar y)  , \label{cat_spec3}\\
&I(\bar x,\bar y)\triangleq\frac{1}{(2\pi)^2}\int\hspace{-0.5em}\int d \bar\xi d\bar\eta\,  e^{i \bar\Phi(\bar\xi,\bar\eta;\bar x,\bar y)}\nonumber\\
&\hspace{3.1em} =3^{-1/6}2^{-4/3} \9{Ai}( \bar u)\9{Ai}(\bar v), \label{I}
\end{align}
with $(\bar u,\bar v)$  given by
\begin{align}
 \big(\bar u,\bar v\big)^\7T= (3/2)^{1/6}\3R^{\7T}_{45^0} \big(3^{-1/2}\bar x,\, \bar y\big)^\7T, \label{uv_bar_sys}
\end{align}
where $\3R$ is given in Eq.~\eqref{rotation}. Equations~\eqref{cat_spec3}--\eqref{uv_bar_sys}  can readily be calculated using standard numerical routines.
Note that in the limit  $\bar\epsilon \rightarrow0$, the series representation in Eq.~\eqref{cat_spec3} reduces to the standard series expansion  in \cite[Eq.~36.8.3]{DLMF}.

\subsection{Numerical Example} \label{num_example}

Figures~\ref{Fig_cat_field} and~\ref{Fig_cat_field2} depict the field intensity at two different frequencies such that $k\tilde\beta=10^4$ and~$10^6$, respectively. The results are shown along the normalized $\bar x$-axis in the $\bar y=0$ and~$\bar y=1$ planes. The observation domain is centered at the point $(\tilde x,z)=(0.006, 0.16)\tilde\beta$ on the edge of caustic~2. The figures compare the field calculated via the canonical integral in Eq.~\eqref{field_cat} (green line), with the PE solution of Eqs.~\eqref{gen_parax_field}--\eqref{parax_field} (blue points), and the GO field of Eq.~\eqref{sdp_field} (red dashed line). The canonical integral has been calculated via the series expansion in Eqs.~\eqref{cat_spec3}--\eqref{I}, keeping only the terms $p=0,...3$ and $m=0,...3$. A decay parameter $\tilde \alpha=10^{-4}$  has been used.

In Fig.~\ref{Fig_cat_field} the canonical integral solution and the PE solution are in an excellent agreement on both the lit and the shadow sides of the caustic. As expected, the GO solution breaks on the caustic, but otherwise it agrees well with the other solutions. Further away on the lit side of the caustic, the GO solution starts to deviate, but in this region it is, in fact, more accurate than the other solutions since it does not rely on the local spectrum approximation or on the PE approximation.

In Fig.~\ref{Fig_cat_field2}, on the other hand, on observes a significant \emph{translational error of the PE solution} (notice, on the  other hand, the perfect agreement between the canonical integral and the GO solutions, except on the caustic). The error of the PE solution is due to the deviation of the PE-based beam trajectory in Eq.~\eqref{beam_axis} from the true ray-based trajectory, and can be parameterized by $\bar x_p$ which is the normalized location of the PE caustic in the $\bar x$-system (recall from Fig.~\ref{Fig_caustic_coord} that the $\bar x$-system is centered on the edge of caustic~2, see also Fig.~\ref{Fig_caustic_cross_z}). $\bar x_p$  can be found using Eq.~\eqref{beam_axis} in conjunction with Eqs.~\eqref{caustic_coord_sys}, \eqref{edge1_coord}--\eqref{edge2_coord} and  \eqref{caustic_coord}. Here we consider only the small $\theta_z$ approximations of $\bar x_p$, obtaining  \begin{align}
 \bar x_p\approx -3^{1/3}(k\tilde\beta)^{2/3}\theta_z^4. \label{x_p_bar}
\end{align}
In Fig.~\ref{Fig_cat_field} the frequency is relatively low such that $\bar x_p=-2.7\times 10^{-2}$ (i.e., $\hat x_p=-6\times 10^{-2}\lambda$), hence the error of the PE solution is hardly noticeable on the figure scale. In Fig.~\ref{Fig_cat_field2}, on the other hand, the frequency is higher giving $\bar x_p=-0.59$ (i.e.,  $\hat x_p=-6\lambda$). Since $\bar x_p$ is $O(1)$, the translational error of the PE solution is readily discerned in the figure.

In addition to the translational error of the PE solution, the exact canonical integral has some intricate features which are readily discerned in the main-lobe of  Fig.~\ref{Fig_cat_field2} (notice the peak $1\%$ error of the PE-solution there). Recalling the discussion after Eq.~\eqref{caustic_param}, the field structure of the canonical integral is parameterized by $\bar \delta$ which is the normalized distance between caustics~1 and~2. As noted there, this parameter grows with both frequency and observation range. For $\bar \delta\ll1$ (near the aperture or small $k\beta$), the distance between these caustics is small and the canonical integral reduced to the Ai function field structure in Eqs.~\eqref{gen_parax_field}--\eqref{parax_field}, but for larger $\bar\delta$, the field structure of the canonical integral describes the separation between the two caustics and is more complicated than the PE solution. This is seen in Figs.~\ref{Fig_cat_field} and~\ref{Fig_cat_field2}  where $\bar\delta=7.6\times10^{-3}$ and $3.5\times10^{-2}$, respectively. In these figures, the cusped edge of caustic~1 is located at $\bar x_\71=-1.7 \times10^{-4}$ and $\bar x_\71=-3.7 \times10^{-3}$, respectively (i.e., $\hat x_\71=-4\times 10^{-4}\lambda$ and $\hat x_\71=-4\times 10^{-2} \lambda$).

Finally, to help the reader we map the parameters used in these numerical examples to the conventional ones used for example in \cite{Broky08,Sivil_Op_let07} (see also discussion after Eq.~\eqref{aperture_field}). Noting that the ``Fresnel distance''  $z_\70=kx_\70^2$ used there is related to our $\tilde\beta$ parameter via $kz_\70=2^{1/3}(k\tilde\beta)^{2/3}$, it  follows that the range at the caustics in Figs.~\ref{Fig_cat_field} and~\ref{Fig_cat_field2} is $z=2.7z_\70$ and $z=12.5z_\70$, respectively.

\section{ Discussion on the Complete Caustic Topology} \label{caustic_complete}

We consider the \emph{global} caustic topology which includes far away parts of the caustic which were not included in the  analysis  of Sec.~\ref{caustic_top_parax}. The field in these regions is  formed by contributions of rays that emanate sideways from far away points in the aperture. This  global picture is presented here in connection with the discussion in  Sec.~\ref{cat_top_dis} regarding the limitations  of the AiB. Since the caustic structure is rather complex,  we discuss separately the topologies of caustic~2 and~1 in Secs.~\ref{caustic_comp2} and~\ref{caustic_comp1}, respectively.

\subsection{Global Structure of Caustic 2} \label{caustic_comp2}

The global structure of caustic~2 is depicted in Fig.~\ref{Fig_caustic3}, as a zoom-out view of Fig.~\ref{Fig_caustic_exact}. In addition to the  sheet shown in Fig.~\ref{Fig_caustic_exact}, tagged here  ``caustics  2A'', there is another caustic surface which is tagged ``caustic~2B.'' These two surfaces  are connected and form a pyramid-like structure which terminates  at a sharp edge at the top. Caustic 2B  is formed by  rays exiting the aperture at far away regions, as illustrated in Fig.~\ref{Fig_caustic_cross_sym} which shows a cross section of the caustic surface in the $\tilde y=0$ (the cross sectional line is also depicted in Fig.~\ref{Fig_caustic3} by  a red dashed-line).

\subsection{Global Structure of Caustic 1} \label{caustic_comp1}

Figs.~\ref{Fig_caustic4} and~\ref{Fig_caustic5} depict the complete structure of caustic~1 as a zoom-out view of Fig.~\ref{Fig_caustic_exact}, with Fig.~\ref{Fig_caustic4} showing the caustic parts formed by ray species $s=1$ (see Fig.~\ref{Fig_exit_angles} for the species definition), and Fig.~\ref{Fig_caustic5} showing the rest of the caustic. The surfaces in Figs.~\ref{Fig_caustic4} and~\ref{Fig_caustic5} are connected at the margins that are delineated in the figures. The connections of these surfaces are clarified in Fig.~\ref{Fig_cross_hat} which depicts  cross-sectional views of the complete caustic at two constant-$z$ planes. A cross sectional view  in the symmetry plane $\tilde y=0$ is depicted in  Fig.~\ref{Fig_caustic_cross_sym}.

As can be discerned from  Fig.~\ref{Fig_caustic4}, the part of caustic~1 that is shown in Fig.~\ref{Fig_caustic_exact}, tagged here as ``caustics  1A'', is connected to a hat-like  structure, tagged  as ``caustic~1B.'' The latter is formed by rays of species~1 exiting the aperture from far away points, as in the case of caustic~2B of Fig.~\ref{Fig_caustic3}. Caustic~1B is  a cuspoid (as may also be seen in Fig.~\ref{Fig_cross_hat}) with its edge in the $\tilde y=0$ plane (red dashed line in Fig.~\ref{Fig_caustic4}).

\subsection{Discussion: Limitations of the AiB Solution} \label{cat_top_dis}

As discussed in Sec.~\ref{caustic_top_parax}, the AiB has been identified as  a hyperbolic umbilic diffraction catastrophe.  However, as can be seen from Figs.~\ref{Fig_caustic3}-\ref{Fig_caustic4}, near the termination point of caustic 2, the  hyperbolic umbilic caustic is no longer isolated, and the caustic structure evolves into the more complex parabolic umbilic catastrophe  \cite[Fig.~7.3]{nye1999natural}. From this global analysis  it then follows that the propagation range of the AiB is limited  to the termination point of caustics~2A (see also Fig.~\ref{Fig_caustic_cross_sym}) where  the  AiB loses its beam shape. The maximal propagation range is of order $\tilde\beta$ and hence it can be controlled by choosing $\tilde\beta$  for a desired application. The range is also controlled by $\tilde\alpha$ which controls the decay rate along the beam axis (see Eqs.~\eqref{field_cat_finite} and \eqref{tau_v_alpha}).

The exact ray-based solutions of Secs.~\ref{field_calc} and~\ref{catastrophe} may be used up to the maximal propagation range discussed above. In the near zone, one may also use the PE solution in Eqs.~\eqref{gen_parax_field}--\eqref{parax_field}, but the accuracy of this solution reduces as the distance between the exact and the PE approximated caustics grows (see Fig.~\ref{Fig_edge_parax_vs_exact}), giving rise to the translational error of the PE field in Fig.~\ref{Fig_cat_field2}.

\subsection{Comments on the Range Limitations of the 2D AiB}
In  \cite{KaganHeyman} and\cite{KaganHeymanJOSA2011} it has been shown that the 2D-AiB is supported by a smooth caustic which is a fold catastrophe \cite{Kravtsov_catastrophe}. Those references, however, have concentrated on the field near the beam axis. A global analysis reveals that  this 2D caustic is actually  a cusp  whose form is very similar to the cross section of caustic~2 in Fig.~\ref{Fig_caustic_cross_sym} (caustic~1 does not exist in the 2D case).
This global structure was not discussed in \cite{KaganHeyman} and \cite{KaganHeymanJOSA2011} since the rays forming the missing parts have weak contributions and late arrival times beyond the pertinent time window  \cite{KaganHeymanJOSA2011}.  The analysis in   Sec.~\ref{cat_top_dis} regarding the limitations  of the AiB can be similarly applied to the 2D case.

\section{Conclusions} \label{con}

We presented a wave-theoretic analysis of the 3D-AiB and derived uniform solutions for the transition layer near the propagation axis. The formulation is based on  exact spectral and ray representations and therefore does not suffer from the limitations of the conventional PE approximation. As in the 2D-AiB case, the 3D-AiB is described by sideways radiating rays that focus onto a caustic that delineates the beam trajectory. However, the 3D ray  topology is considerably more complicated than the 2D topology, and it has many intricate features. Specifically, near the curved beam axis, the ray topology of the 3D-AiB consists of a cusped double-layered caustic identified as the hyperbolic umbilic catastrophe (Sec.~\ref{caustic_top}).  At a certain range away from the aperture, the hyperbolic umbilic  caustic  is no longer isolated and evolves into the more complex parabolic umbilic caustic, where the AiB loses its beam shape.

To clarify these issues we have investigated the ray topology of the 3D-AiB using the exact spectral representation for the AiB. We concentrated first on the region near the curved beam axis (Sec.~\ref{caustic_top}), but we have also explored the \emph{global} caustic topology (Sec.~\ref{caustic_complete}). Specifically, we have thoroughly explored all the ray paths that form the double-layered hyperbolic umbilic caustic noted above (Figs.~\ref{Fig_caustic1_ray_skeleton}--\ref{Fig_caustic_coord}). These contributions are essential in order to understand the time-dependent Airy pulsed beam (AiPB) that will be presented in a subsequent publication \cite{KaganHeyman_3DAiB_part2_JOSA}.   

The field calculations in Secs.~\ref{field_calc}--\ref{catastrophe} were based on an asymptotic evaluation of the exact spectral integral of the AiB. In order to calibrate the final results, we derived first the lowest order geometrical optics (GO) solution (Sec.~\ref{GO}) which, however, fails near the beam axis. The field near the beam axis has been considered separately in Sec.~\ref{catastrophe} within the framework of  catastrophe theory \cite{poston1996catastrophe,Kravtsov_catastrophe,nye1999natural}. As discussed in the introduction (see also Sec.~\ref{catastrophe}),  a  \emph{globally} uniform field solution can be obtained via the  procedure in \cite[Sec.~5]{berry1976waves}. This approach, while very elegant and systematic, leads to \emph{implicit} representations where the field is described in a generalized coordinate system that has to be calculated numerically. We therefore used in Sec.~\ref{cat_type} an alterative \emph{local} approach where the spectral integral of the AiB has been approximated  near the caustic using a local coordinate system, and the final result has been expressed as a canonical hyperbolic-umbilic diffraction integral \cite[Sec.~36.2]{DLMF}. This more direct approach does not have the global uniformity of the general implicit procedure in \cite{berry1976waves}, yet it has led to a solution that is expressed explicitly in the local coordinate system of the caustic, where all the field parameters are expressed in terms of the geometrical parameters of the caustic and  of the aperture field distribution. This solution is valid in the transition layer near the caustic and reduces correctly to the GO field away from the caustic, but far away, one has to switch to the GO field of Sec.~\ref{GO} noted above. Using this approach we also modified the hyperbolic-umbilic integral by introducing a complex loss parameter that describes the finite energy AiB.

The final field solution in Eqs.~\eqref{field_cat_finite} with \eqref{cat_spec} and \eqref{phi_canon_finite}) provides an \emph{explicit} solution in the transition layer near the beam axis.  The width of the transition (or boundary) layer is of order $k^{-2/3}\tilde{\beta}^{1/3}$ (see $\bar x$, $\bar y$ coordinates in Eq.~\eqref{caustic_coord}) where $k$ is the wave-number and $\tilde\beta$ is a frequency-independent parameter describing  the beam geometry (see Eq.~\eqref{beam_axis}). The caustic topology is controlled by the dimensionless parameter $\bar\delta$ of Eq.~\eqref{caustic_param} appearing in $\Phi$ of Eq.~\eqref{phi_canon_finite}. This parameter represents the normalized distance between the two layers of the hyperbolic umbilic caustic, scaled with the normalized frequency like $(k\tilde{\beta})^{1/3}$. As $\bar\delta$ grows with the observation range $z$, the structure of the boundary layer field described by the canonical integral becomes  more complex while for $\bar\delta \gg 1$, the two layers are separable and the field can be described by separate cusp and fold catastrophes, as discussed after Eq.~\eqref{caustic_param}.

The canonical hyperbolic umbilic integral can be evaluated efficiently via contour deformation in the complex plane. Here, however, we calculated it using the series expansion of \cite[Sec.~36.8]{DLMF} which has been modified to accommodate the finite energy AiB (Sec.~\ref{cat_field}). We used it to calculate the field near the beam axis under several parameter regimes (Sec.~\ref{num_example}). The accuracy of the canonical integral has been demonstrated by noting in Fig.~\ref{Fig_cat_field} that away from the beam axis it reduces to the GO field (the GO field near the hyperbolic umbilic caustic is a sum of four rays).  Far away from the  transition layer, however, the field is described more accurately by the GO solution of Sec.~\ref{GO}, since it is structured upon the \emph{exact} ray skeleton and not on the \emph{local} ray skeleton near the beam axis used here for the construction of the transition function (note that the solutions obtained via the \emph{globally} uniform approach in \cite[Sec.~5]{berry1976waves} reduce correctly to the GO field even very far from the transition layer, yet, as noted earlier, these solutions are implicit functions of the space coordinates). Comparing the canonical integral to the conventional PE solution one observes in Fig.~\ref{Fig_cat_field2} a significant translational error of the latter which is due to the error of the PE approximated beam trajectory. This error is parameterized by the dimensionless parameter $\bar x_p$ of Eq.~\eqref{x_p_bar} which describes the distance between the exact and the PE caustics. $\bar x_p$  grows with $z$ (see  Figs.~\ref{Fig_caustic_cross_z} and~\ref{Fig_edge_parax_vs_exact}) and scales with the normalized frequency like  $(k\tilde{\beta})^{2/3}$. As discerned from Figs.~\ref{Fig_cat_field}--\ref{Fig_cat_field2}, the error is small for $\bar x_p\ll1$, but it becomes significant for $\bar x_p\sim 1$.

In summary, it has been demonstrated that the local canonical integral in conjunction with GO provides an accurate and convenient solution to the AiB, as an alternative to the conventional PE solution. This solution is structured upon the exact ray skeleton and therefore has a wider validity range than the PE solution. Furthermore, this solution provides a convenient design tool for synthesizing other types of beams propagating along  any prescribed  convex trajectory in uniform or non uniform medium. An application of this idea in 2D configurations has been presented in \cite{AiB_arbitrary_traj_PRL2011}.

The analysis in this work has been performed within the framework of the \emph{non-dispersive} AiB where the aperture field is scaled with frequency such that the ray skeleton is frequency-independent. We use the non-dispersive formulation because it simplifies the ray analysis by making it more transparent geometrically. Moreover, this formulation also extends the AiB solution to the ultra wide band (UWB) frequency regime by ensuring that the pulsed field propagates along the curved beam trajectory without dispersion. This non-dispersive formulation will be used in a subsequent publication \cite{KaganHeyman_3DAiB_part2_JOSA} to introduce a new class of 3D Airy pulsed beams (AiPB).

% \section*{Acknowledgment}
% This work is supported in part by the Israeli Science Foundation, under Grant No. 674/07 and 263/11.
% {\R The authors would  like to thank the reviewers for  their helpful and constructive comments.}

\renewcommand{\thesection}{\Alph{section}}
\setcounter{section}{0}
\renewcommand{\thesubsection}{\Alph{section}.\arabic{subsection}}
\renewcommand{\thesubsubsection}{\Alph{section}.\arabic{subsection}.\arabic{subsubsection}}
\renewcommand{\theequation}{\Alph{section}.\arabic{equation}}
\setcounter{equation}{0}
\section{Appendix A: Derivation of the GO Field from the Spectral Representation } \label{exact_caustic_eqs}

In Sec.~\ref{caustic_top} we introduced the ray representation of the AiB based on the \emph{approximate} initial field distribution in Eq.~\eqref{IC_for_rays}. In this appendix we derive the ray representation based on the \emph{exact} spectral representation of the initial field distribution in Eq.~\eqref{beam_freq}, and show that the results are the same as those of Sec.~\ref{caustic_top}.

As discussed in Eq.~\eqref{IC_for_rays}, $\alpha$ is relatively small and hence we consider first the ray and caustic equations for $\alpha=0$ (Secs.~\ref{ray_eqs_appx}--\ref{calc_parax_caustic}) and then we consider the  GO field for the case $\alpha\neq 0$ (Sec.~\ref{go_field_appx}).

\subsection{The Ray Equation} \label{ray_eqs_appx}
We start by deriving the ray equation in Eq.~\eqref{ray_coord} from the saddle point condition of the exact spectral representation in Eq.~\eqref{beam_freq}.

Substituting the spectral delay function $\tau$ of  Eq.~\eqref{tau} with $\alpha=0$ into the saddle points condition in Eq.~\eqref{stationary_points} we obtain
\begin{align}
&\partial_\xi\tau=\partial_\xi{\tau_\70}+x/c-(z/c)(\xi/\zeta)=0, \label{d_xi_tau_gen} \\
&\partial_\eta \tau=\partial_\eta{\tau_\70}+y/c-(z/c)(\eta/\zeta)=0,\label{d_eta_tau_gen}
\end{align}
where $\zeta$ is defined in Eq.~\eqref{zeta}. For $z=0$, Eqs.~\eqref{d_xi_tau_gen}--\eqref{d_eta_tau_gen} reduce to
\begin{align}
&\partial_\xi{\tau_\70}=-x'/c=\beta\xi^2/c, \label{d_tau0_gen1} \\
&\partial_\eta{\tau_\70}=-y'/c=\beta\eta^2/c,\label{d_tau0_gen2}
\end{align}
where $(x',y')$ are the coordinates in the $z=0$ plan. The right hand side in Eqs.~\eqref{d_tau0_gen1}--\eqref{d_tau0_gen2} is obtained by using Eq.~\eqref{tau_0} for $\tau_\70$  and it is identical to Eq.~\eqref{exit_angles} for the ray exit angles. Substituting the left hand side of Eqs.~\eqref{d_tau0_gen1}--\eqref{d_tau0_gen2} back into Eq.~\eqref{d_xi_tau_gen}--\eqref{d_eta_tau_gen} and rearranging, we obtain
\begin{align}
&x=x'+\xi (z/\zeta), \label{ray_eq_app1} \\
&y=y'+\eta (z/\zeta), \label{ray_eq_app2},
\end{align}
which are, in fact  the expressions for the ray trajectories in Eq.~\eqref{ray_coord} with $\sigma=z/\zeta$.  It follows that the ray skeleton derived in Sec.~\ref{caustic_top} using the approximation  in Eq.~\eqref{Airy_asymp} is the same as the one derived here from the exact spectral representation.

\subsection{Ray and Caustic Topology - Exact Analysis} \label{calc_exact_caustic}
There are two alternative routs to derive the expression for the caustic of  Sec.~\ref{caustic_top}. The first is to calculate $\cal{J}$ of  Eq.~\eqref{Jacobian} using the ray trajectories in Eq.~\eqref{ray_coord}, and then to solve for its zeros. The second relies on the spectral formulation of Sec.\ref{spec}. Here we start with the latter since it also complies with the formalism of catastrophe theory which is discussed in Sec.~\ref{catastrophe}.

To calculate the caustic surface we note that the elements of the Hessian matrix $\3H$ in Eq.~\eqref{Hessian} are given by
\begin{align}
&(c/\beta)\partial^2_\xi\tau=2\xi-(z/\beta)(1-\eta^2)/\zeta^3, \label{tau_xi_xi} \\
&(c/\beta)\partial^2_\eta\tau=2\eta-(z/\beta)(1-\xi^2)/\zeta^3,   \label{tau_eta_eta}\\
&(c/\beta)\partial_\eta\partial_\xi\tau=(c/\beta)\partial_\xi\partial_\eta\tau=-(z/\beta)\xi\eta/\zeta^3,   \label{tau_xi_eta}
\end{align}
and hence, the  determinant $H$ is given by
\begin{align}
   &H=[C\71 z^2+ C_\72\beta z+ C_\73\beta^2]/c^2, \label{H_eq2}\\
&C_\71(\xi,\eta)=1/\zeta^4, \qquad C_\73(\xi,\eta)=4\xi\eta, \label{c1}   \\
&C_\72(\xi,\eta)=-2[\xi(1-\xi^2)+\eta(1-\eta^2)]/\zeta^3. \label{c2}
\end{align}

At the caustic, two or more rays (saddle points)  coalesce, implying that $H=0$ there. This equation should be solved in conjunction with Eqs.~\eqref{d_xi_tau_gen}--\eqref{d_tau0_gen2} since the caustic is composed of rays.  From Eq.~\eqref{H_eq2} a ray with  parameters $(\xi,\eta)$  is tangent to the  caustic  at two points
\begin{align}
  z_{c\7{1,2}}/\beta=(-C_\72\pm\sqrt{C_\72^2-4C_\71C_\73})/2C_\71.  \label{zc12}
\end{align}
The $(x,y)$ coordinates of these points are found by substituting Eq.~\eqref{zc12} back into  Eqs.~\eqref{d_xi_tau_gen}--\eqref{d_eta_tau_gen}.

As shown in Sec.~\ref{caustic_top} and Fig.~\ref{Fig_caustic_exact}, the exact caustic topology near the beam axis is composed of two sheets, denotes as caustic~1 and caustic~2. We have also shown in Figs.~\ref{Fig_caustic1_ray_skeleton}--\ref{Fig_caustic2_ray_skeleton} that a ray of species~1 touches caustic~2 first and then touches caustic~1, never touching the caustic again. Thus the ``$\pm$'' sign  in Eq.~\eqref{zc12}  corresponds to a point on  caustic~1 and~2, respectively.

The edges of the caustics in the symmetry plane $\tilde y =0$ (see Figs.~\ref{Fig_caustic_exact} and~\ref{Fig_caustic_cross_z}) are found by setting  $\xi=\eta$ in Eqs.~\eqref{zc12} and \eqref{d_xi_tau_gen}--\eqref{d_eta_tau_gen}. The $(\tilde x,  z$) coordinates can be expressed in the following parametric form
\begin{align}
  &(\tilde x_\71,z_\71)\triangleq\tilde\beta(\sin^2\theta_z,\sin2\theta_z),   \label{edge1_coord} \\
  &(\tilde x_\72,z_\72)\triangleq\tilde\beta(\sin^2\theta_z\cos2\theta_z,\sin2\theta_z\cos^2\theta_z),  \label{edge2_coord}
\end{align}
where $\sin\theta_z=(\xi^2+\eta^2)^{1/2}$, $\theta_z$ being the ray-angle with respect to the $z$ axis (see Eq.~\eqref{ray_direct}). Alternatively, $z_\7{1,2}$ in Eqs.~\eqref{edge1_coord}--\eqref{edge2_coord} can be expressed as function of $\tilde x$
\begin{align}
 & z_\71/\tilde\beta=2(\tilde x/\tilde\beta)^{1/2}(1-\tilde x/\tilde\beta)^{1/2}, \label{edge1_eq} \\
 & z_\72/\tilde\beta=2q^{1/2}(1-q)^{3/2}, \qquad q=(1-(1-8\tilde x/\tilde\beta)^{1/2})/4, \label{edge2_eq}
\end{align}
where all square roots taken with a positive real part.
At moderate ranges such that $\tilde x/\beta\ll1$ these expressions may be approximated by
\begin{align}
 & z_\71/\tilde\beta \approx 2(\tilde x/\tilde\beta)^{1/2}(1-\tilde x/4\tilde\beta), \label{vertex1_approx} \\
 & z_\72/\tilde\beta \approx 2(\tilde x/\tilde\beta)^{1/2}(1-3\tilde x/2\tilde\beta), \label{vertex2_approx}
\end{align}
where we used $(1-x)^{n}\approx 1-nx$. The lowest order approximation in Eqs.~\eqref{vertex1_approx}--\eqref{vertex2_approx} is the PE solution $z_p/\tilde\beta \approx 2(\tilde x/\tilde\beta)^{1/2}$ of Eq.~\eqref{beam_axis}, hence one finds for a given $\tilde x$ that $z_\72<z_\71<z_p$ (see also Fig.~\ref{Fig_edge_parax_vs_exact}).

Finally we note that the same results could have been obtained via the ray-system analysis. The Jacobian in Eq.~\eqref{J_app} can be expressed explicitly as
\begin{align}
 \5J(\sigma)=C_\71\zeta^3\sigma^2+\beta C_\72\zeta^2\sigma+\beta^2C_\73\zeta , \label{J_exc}
\end{align}
where $C_\7{1,2,3}$ are given in Eqs.~\eqref{c1}--\eqref{c2}.  Comparing Eqs.~\eqref{H_eq2} and~\eqref{J_exc}, it follows that the zeros $\sigma_{c\7{1,2}}$ of $\5J(\sigma)$ correspond to the zeros $z_{c\7{1,2}}$ of $H$ in Eq.~\eqref{zc12}.  $\sigma_{c\7{1,2}}=z_{c\7{1,2}}/\zeta$ define the ray lengths to the caustic.

\subsection{The Caustic within the Parabolic Equation Approximation} \label{calc_parax_caustic}
In the paraxial regime  one can approximate in Eq.~\eqref{zeta} $\zeta\approx1-\xi^2/2-\eta^2/2$.  Substituting $\zeta\approx 1$ in the ray equations \eqref{d_xi_tau_gen}--\eqref{d_eta_tau_gen} with \eqref{d_tau0_gen1}--\eqref{d_tau0_gen2}, yields
\begin{align}
 &\xi=z/2\beta\pm\sqrt{(z/2\beta)^2-x/\beta}, \label{xi_ray_parax} \\
 &\eta=z/2\beta\pm\sqrt{(z/2\beta)^2-y/\beta}. \label{eta_ray_parax}
\end{align}
The solutions of Eq.~\eqref{H_eq2} yields two surfaces $S_\7{1,2}$, given by
\begin{align}
  &S_\71: (z/2\beta)^2-x/\beta=0 \quad\& \quad y/\beta \geq (z/2\beta)^2, \label{caustic_S1} \\
  &S_\72: (z/2\beta)^2-y/\beta=0 \quad \& \quad x/\beta \geq (z/2\beta)^2, \label{caustic_S2}
\end{align}
which are shown in Fig.~\ref{Fig_caustic_parax}. These surfaces intersect at the edge which is the beam axis in Eq.~\eqref{beam_axis}.

\subsection{Details of the Derivation of the GO Field in Eqs.~\eqref{sdp_field}--\eqref{H_i}} \label{go_field_appx}

The phase term $\tau_r$ in Eq.~\eqref{tau_i} is obtained by substituting the ray directions in Eqs.~\eqref{d_tau0_gen1}--\eqref{d_tau0_gen2} (see also Eq.~\eqref{exit_angles}) and the ray equation in Eq.~\eqref{ray_direct} into $\tau(\xi,\eta)$ of  Eq.~\eqref{tau}.

Next, in order to derive the amplitude term in Eq.~\eqref{H_i}, we note from Eq.~\eqref{H_eq2} and Eq.~\eqref{J_exc} that the  determinant of the Hessian and the Jacobian can be cast in the form
 \begin{align}
 H(\sigma)= \5J(\sigma)/\zeta c^2= (\sigma-\sigma_{c\71})(\sigma-\sigma_{c\72})/c^{2}, \label{H_eq3}
 \end{align}
where $\sigma= z/\zeta$ and $\sigma_{c\7{1,2}}$ correspond to $z_{c\7{1,2}}$ of Eq.~\eqref{zc12}; $\sigma_{c\7{1,2}}$  are the ray lengths to caustic~1 and~2, respectively.

From Eq.~\eqref{H_eq3}, the eigenvalues of the Hessian matrix defined in Eq.~\eqref{eigen_H} are given by
  \begin{align}
 h_\7{1,2}=(\sigma-\sigma_{c\7{1,2}})/c.
 \label{d12}
 \end{align}
Thus $h_\71<0$ if the ray has not touched caustic~1 yet, while $h_\71>0$ if it has touched it. Similarly, $h_\72\lessgtr 0$ if the ray has not yet touched or has touched caustic~2, respectively. We also note, by analyzing Eq.~\eqref{zc12} for $z_{c\7{1,2}}$ using $C_\7{1,2,3}$ from Eqs.~\eqref{c1}--\eqref{c2}, that the ray species $s=1$ has $\sigma_{c\7{1,2}}\geq0$; species $s=2,3$ have $\sigma_{c\72}\leq0$ and $\sigma_{c\71}\geq0$; and species $s=4$ has  $\sigma_{c\7{1,2}}\leq0$. Combining these results we find that $\mu_r$ of  Eq.~\eqref{eigen_H} can be written as
\begin{align}
 \mu_r=\9{sgn}\{h_{\71r}\}+\9{sgn}\{h_{\72r}\}=2(-1+M_{\70s}+M_r), \label{eigen_H_app}
\end{align}
where $M_{\70s}$ is specified after Eq.~\eqref{A_0s} and $M_r$ is the Maslov index defined after Eq.~\eqref{H_i} that counts the number of times that the $r$th ray has touched a caustic.

Next, using the left hand side in Eq.~\eqref{H_eq3} in conjunction with Eqs.~\eqref{d_tau0_gen1}--\eqref{d_tau0_gen2} we find that $|H|^{-1/2}$ in Eq.~\eqref{sdp_field} can be expressed as
\begin{align}
|H|^{-1/2}=c2^{-1}\beta^{-1/2}(-x')^{-1/4}(-y')^{-1/4} |\5J(0)/\5J(\sigma)|^{1/2}. \label{J_exc2}
\end{align}
The final result for the GO amplitude in the right hand side of Eq.~\eqref{H_i} is obtained by substituting Eqs.~\eqref{eigen_H_app}--\eqref{J_exc2} in the left hand side of that equation.

\subsection{Finite Energy AiB ($\alpha\neq0$)} \label{sec-alpha_neq0}

For $\alpha\neq0$ the saddle points $(\xi,\eta)$ are complex and are given by Eqs.~\eqref{d_xi_tau_gen}--\eqref{d_tau0_gen2} with $\xi$ and $\eta$  in Eqs.~\eqref{d_tau0_gen1}--\eqref{d_tau0_gen2} replaced by $\xi+i\alpha$ and $\eta+i\alpha$, respectively. Since $\alpha$ is small, the solution for the complex saddle points is very close to the real saddle points $(\xi_r,\eta_r)$ obtained for $\alpha=0$.  Therefore, we expand $\zeta$ in Eq.~\eqref{d_xi_tau_gen}-\eqref{d_eta_tau_gen} to first order about $(\xi_r,\eta_r)$
\begin{align}
  \zeta \approx \zeta_r +\zeta^\xi_r(\xi-\xi_r)+\zeta^\eta_r(\eta-\eta_r), \label{zeta_approx}
\end{align}
where $\zeta_r=\zeta$, $\zeta^\xi_r=\partial_\xi\zeta$, $\zeta^\eta_r=\partial_\eta\zeta$, all taken at $(\xi,\eta)=(\xi_r,\eta_r)$, and are given by
\begin{align}
\zeta^\xi_r=-\xi_r/\zeta_r, \quad \zeta^\eta_r=-\eta_r/\zeta_r. \label{zeta_xi_eta}
\end{align}
Substituting into
Eqs.~\eqref{d_xi_tau_gen}--\eqref{d_eta_tau_gen} and solving for the saddle points $(\xi,\eta)$ yields
\begin{align}
(\xi,\eta)=(\xi_r,\eta_r)-i(\alpha,\alpha). \label{stat_point_complex}
\end{align}
Substituting Eq.~\eqref{stat_point_complex},  and~\eqref{zeta_approx}--\eqref{zeta_xi_eta} into $\tau$ of Eq.~\eqref{tau}, using Eqs.~\eqref{ray_coord} and~\eqref{d_tau0_gen1}--\eqref{d_tau0_gen2} and neglecting terms $O(\alpha^2)$ and lower, we obtain
\begin{align}
  c\tau_r=c\tau_{\70r}+\sigma_r-i\alpha(x'_r+y'_r). \label{tau_final}
\end{align}
$\tau_r=c\tau_{\70r}+c\sigma_r$ is the expression in Eq.~\eqref{tau_i}.
The term $\alpha(x'_r+y'_r)$ in Eq.~\eqref{tau_final} appears inside the exponent in Eq.~\eqref{A_0s} where   it is interpreted as a factor in the initial amplitude $A_{\70s}$ rather then a complex time delay.
In the expression for the Hessian we neglect $\alpha$ so Eq.~\eqref{J_exc2} is without change.

\section{Appendix B: Additional Canonical Form for the Hyperbolic Umbilic Catastrophe} \label{Alt_canon}
\setcounter{equation}{0}
In order to prevent possible confusion we note that the Hyperbolic Umbilic catastrophe has another canonical form in addition to the one in Eq.~\eqref{phi_canon}, which has the following form \cite[Sec.~36.3,~Eq.~(3)]{DLMF},\cite[pp.~37]{KravtzovOrlov90}
 \begin{align}
 \Phi(\check\xi,\check\eta;\check x,\check y;\check\delta)=\check \xi^3+\check \eta^3+\check \xi\check \eta\check\delta+\check \xi \check x+\check \eta\check  y.
 \end{align}
This form is obtained from Eq.~\eqref{phi_canon} by using the following transformations
\begin{align}
& \big(\check x,\check y  \big)^\7T=2^{\7{-1/6}}\3R^\7T_{45^\circ} \big([\bar x+3\bar \delta^2/4],3^{\7{1/2}}\,\bar y \big)^\7T , \\
&\big(\check \xi,\check \eta\big)^\7T=2^{\7{1/6}}\3R^\7T_{45^\circ}\big([\bar\xi+\bar \delta/2],3^{\7{-1/2}}\,\bar\eta\big)^\7T, \\
&\check\delta=-3\bar\delta/2^{1/3},
\end{align}
where $(\bar x,\bar y)$ and $\bar \delta$ are defined in Eqs.~\eqref{caustic_coord}--\eqref{caustic_param}, $(\bar\xi,\bar\eta)$ are defined in Eq.~\eqref{caustic_spec}, and $\3R$ is given in Eq.~\eqref{rotation}.
Note that some terms which do not depend on $(\bar \xi,\bar \eta)$  have been omitted since they do not affect the catastrophe type.

%\bibliographystyle{osajnl}
%\bibliography{C:/Users/User/Documents/My_bib_files/yan}
\clearpage

\begin{figure}[tb]
\centering
      \includegraphics[width=9cm]{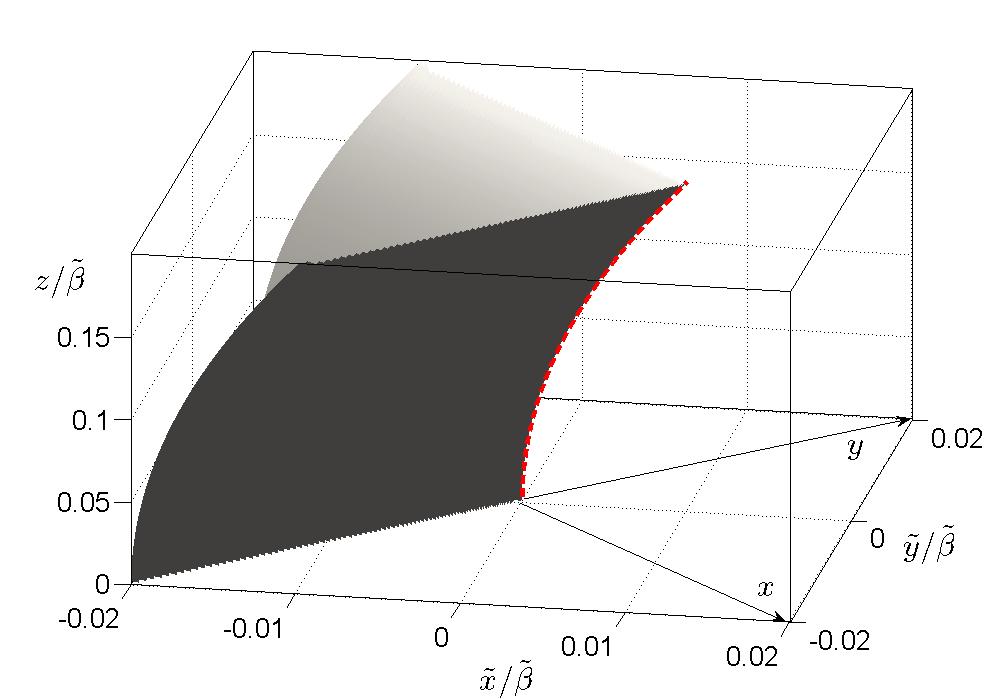}
\caption {(Color online) The caustic of the  AiB within the PE approximation (gray surface). The caustic is analyzed in Appendix~\ref{calc_parax_caustic}. It consists of two curved sheets $S_\7{1,2}$, defined respectively, in Eqs.~\eqref{caustic_S1}--\eqref{caustic_S2} (dark and lit sheets in Fig.~(a), respectively). The sheets are joined at the edge (dashed red line) which delineates the beam propagation trajectory. Cross-sectional cuts are shown in Fig.~\ref{Fig_caustic_cross_z}.} \label{Fig_caustic_parax}
\end{figure}

\begin{figure}[tb]
\centering
     \subfigure[$z/\tilde\beta=0.2$]{%
      \label{Fig_cross1}%
      \includegraphics[width=8cm]{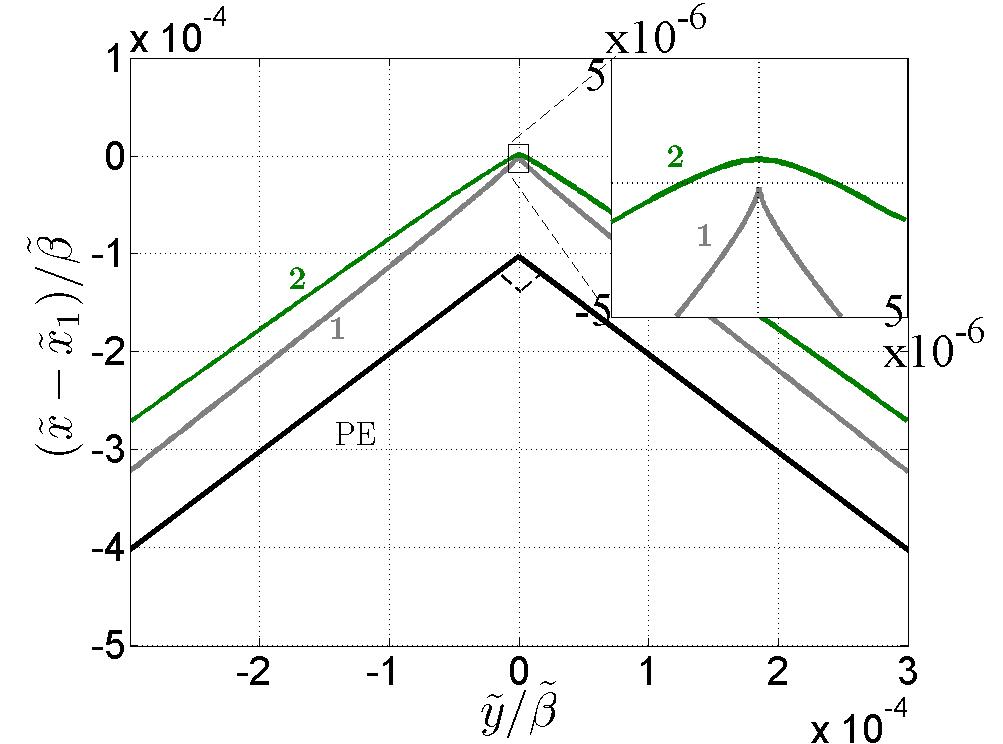}}%
      \\
   \subfigure[$z/\tilde\beta=0.25$]{%
    \label{Fig_cross2}%
    \includegraphics[width=8cm]{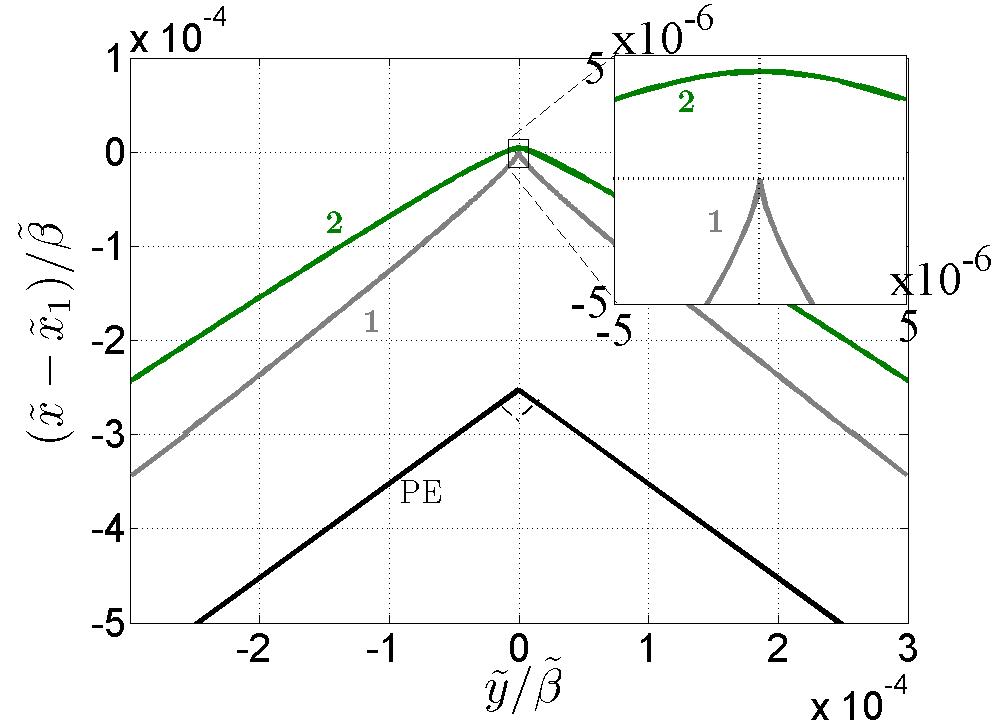}}%
\caption {(Color online)  Cross sections of the  caustics of the AiB at two different ranges:  (a) $z/\tilde\beta=0.2$; and (b) $z/\tilde\beta=0.25$. Black line: the caustic of the PE approximation in Fig.~\ref{Fig_caustic_parax}. Gray and green  lines: the exact caustic, which consists of two surfaces denoted as ``caustic~1'' and ``caustic~2'' respectively (see Sec.~\ref{caustic_top_parax_exact} and Fig.~\ref{Fig_caustic_exact}). The $\tilde x$ axis in this figure is centered about $\tilde x_\71(z)$ the edge of caustic 1 in Eq.~\eqref{edge1_eq}. The insets zoom on the area near the origin where caustic~1  has a ``cusped edge'' while caustic 2  has a smooth corner. One observes that the distance between caustics~1 and~2 and the PE approximated caustic increases with $z$ (see also Fig.~\ref{Fig_edge_parax_vs_exact}).} \label{Fig_caustic_cross_z}
\end{figure}

\begin{figure}[tb]
\centering
    \includegraphics[width=10cm]{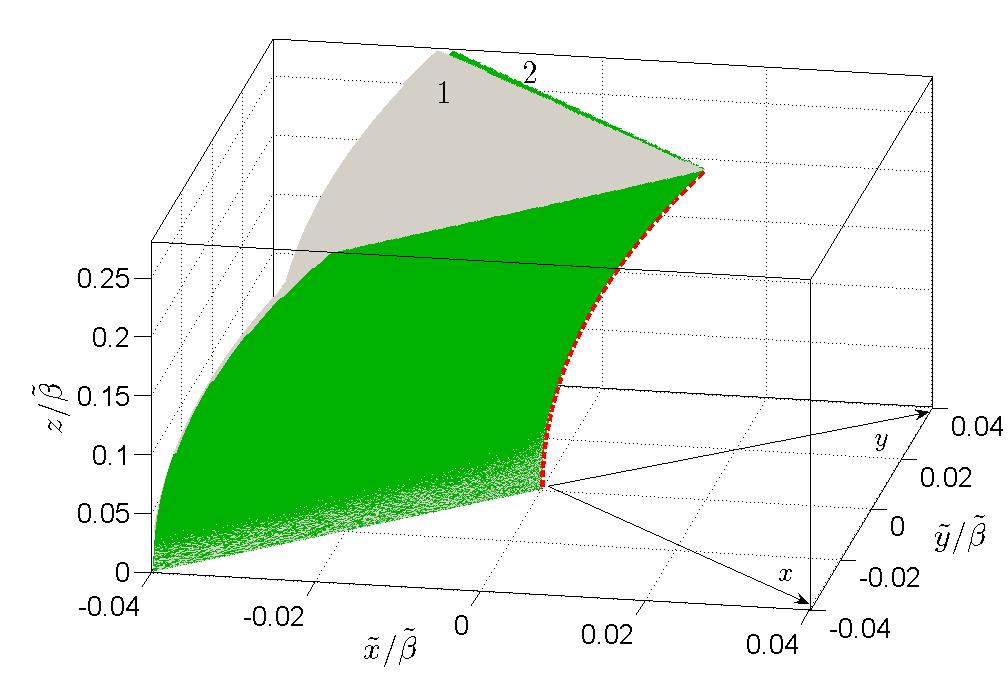}%
\caption {(Color online) The exact caustic of the AiB, discussed in Sec.~\ref{caustic_top_parax_exact} and in Appendix~\ref{calc_exact_caustic}. It consists of two surfaces denoted as ``caustic~1'' and ``caustic~2'' (gray and green surfaces, respectively). The distance between these surfaces is hardly noticeable on the scale of this Figure, but it is shown in the cross-sectional cut of the caustics in Fig.~\ref{Fig_caustic_cross_z}.
Each caustic consists of two surfaces joined at an edge (dashed red lines). The ray formation  of caustics~1 and~2 are  described in Figs.~\ref{Fig_caustic1_ray_skeleton} and~\ref{Fig_caustic2_ray_skeleton}, respectively.} \label{Fig_caustic_exact}
\end{figure}

\begin{figure}[tb]
\centering
    \includegraphics[width=9cm]{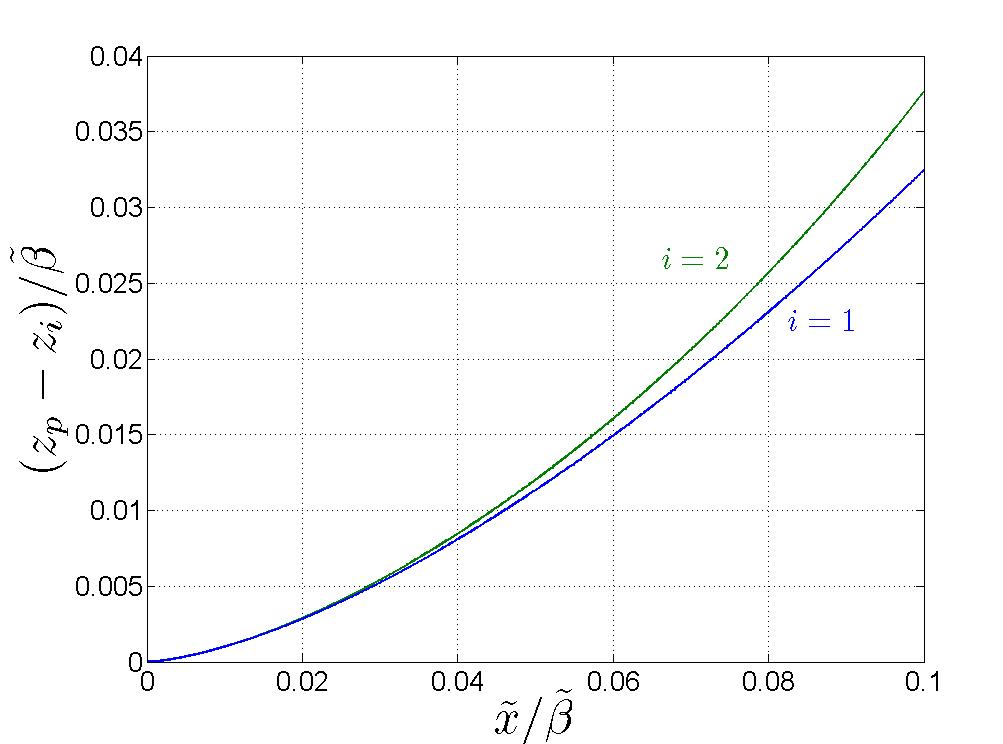}%
\caption {(Color online) The difference between the edges of the PE approximated and of the exact caustics. The figure plots $z_p-z_i$ as a function of $\tilde x$, where $z_p$ is the coordinate of the paraxially approximated beam trajectory in Eq.~\eqref{beam_axis}, and $z_i$, $i=1,2$, are the coordinates of the exact edges of caustic~$i$, given in Eqs.~\eqref{edge1_eq}--\eqref{edge2_eq}. From Eq.~\eqref{beam_axis}, the horizontal range $\tilde x/{\tilde\beta}=0.1$ corresponds to the observation range $z_p/{\tilde\beta}=2\sqrt{0.1}$ and there  $(z_p-z_i)/z_p \approx 5\%-6\%$.} \label{Fig_edge_parax_vs_exact}
\end{figure}

\begin{figure}[tb]
\centering
      \includegraphics[width=12cm]{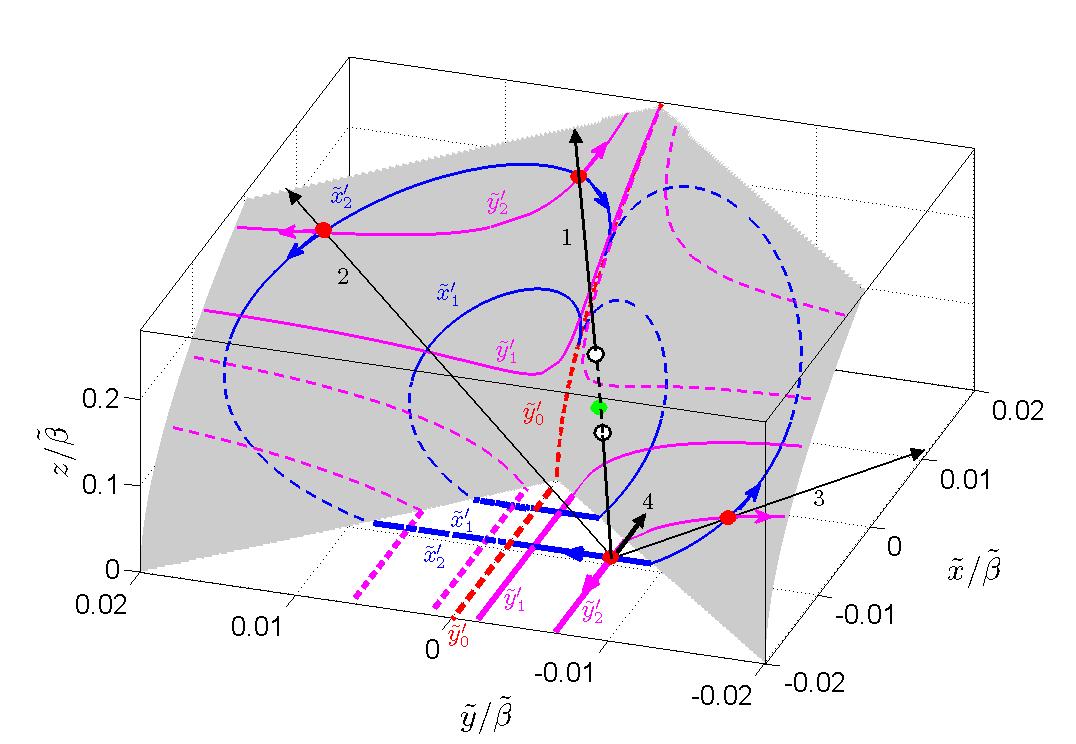}%
\caption {(Color online)  Ray formation of caustic~1 (gray surface) of Fig.~\ref{Fig_caustic_exact}. Each exit point $(\tilde x',\tilde y')$ in the aperture (red point)  emits  four rays (black lines) belonging to species $s=1$,...4 of Fig.~\ref{Fig_exit_angles}. Ray~4 does not touch the caustic. Rays~2,3 touch only caustic~1 (red points). Ray~1 penetrates caustic~1 (through the white circle), touches caustic~2 (green point), penetrates caustic~1 again (through the second white circle) and finally touches caustic~1 (red point), never touching the caustic again. The figure shows  traces of  the exit points in the aperture (thick lines) and the corresponding traces of the points of tangency on the caustic (thin lines with the arrows indicating the  direction of the traces). To emphasize the symmetry, traces with $\tilde y'\gtrless0$ appear as full or dashed lines. Lines of constant $\tilde y'$ and of constant $\tilde x'$ are plotted in magenta and blue, respectively. The traces corresponding to the edge of caustic~1 are shown as red dashed-lines. The parameters used here are: $\tilde x'_\71=-0.004\tilde\beta$, $\tilde x'_\72=-0.009\tilde\beta$, $\tilde y'_\71=\mp0.0015\tilde\beta $, $\tilde y'_\72=\mp0.0065\tilde\beta$, $\tilde y'_\70=0$.}
\label{Fig_caustic1_ray_skeleton}
\end{figure}

\begin{figure}[tb]
\centering
      \includegraphics[width=12cm]{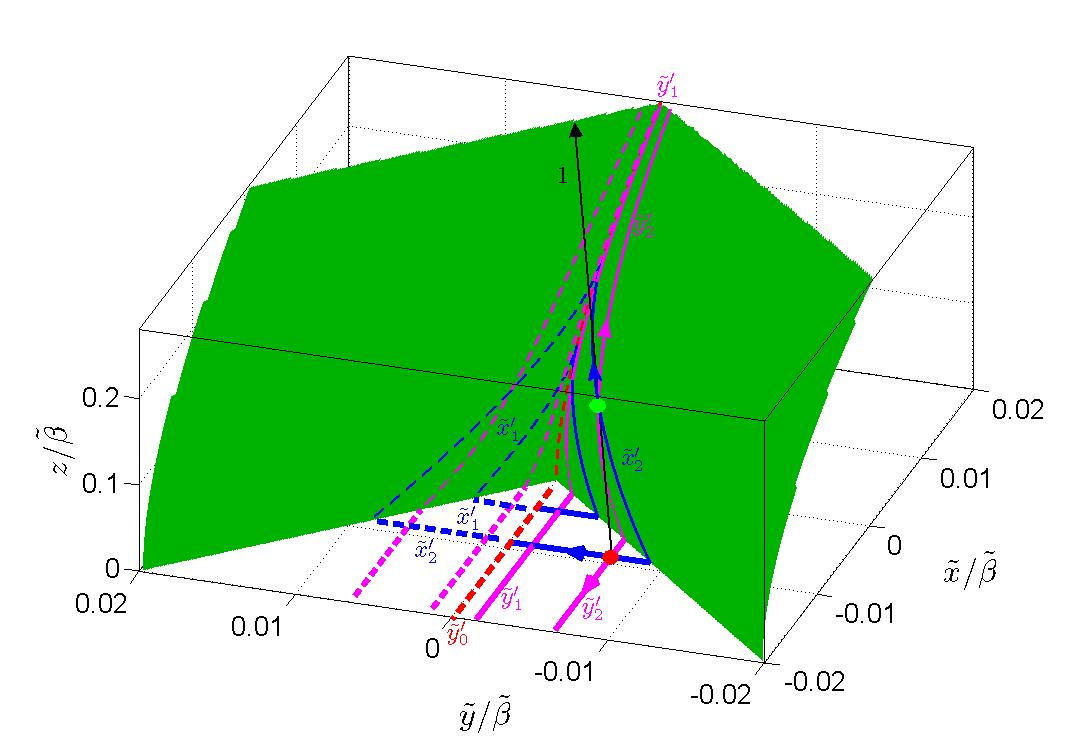}%
\caption {(Color online) Ray formation of caustics~2 (green surface) of Fig.~\ref{Fig_caustic_exact}. The notations are the same as in Fig.~\ref{Fig_caustic1_ray_skeleton}. Note that caustic~2 is formed only by ray species~1. }
\label{Fig_caustic2_ray_skeleton}
\end{figure}

 \begin{figure}[tb]
\centering
      \includegraphics[width=11cm]{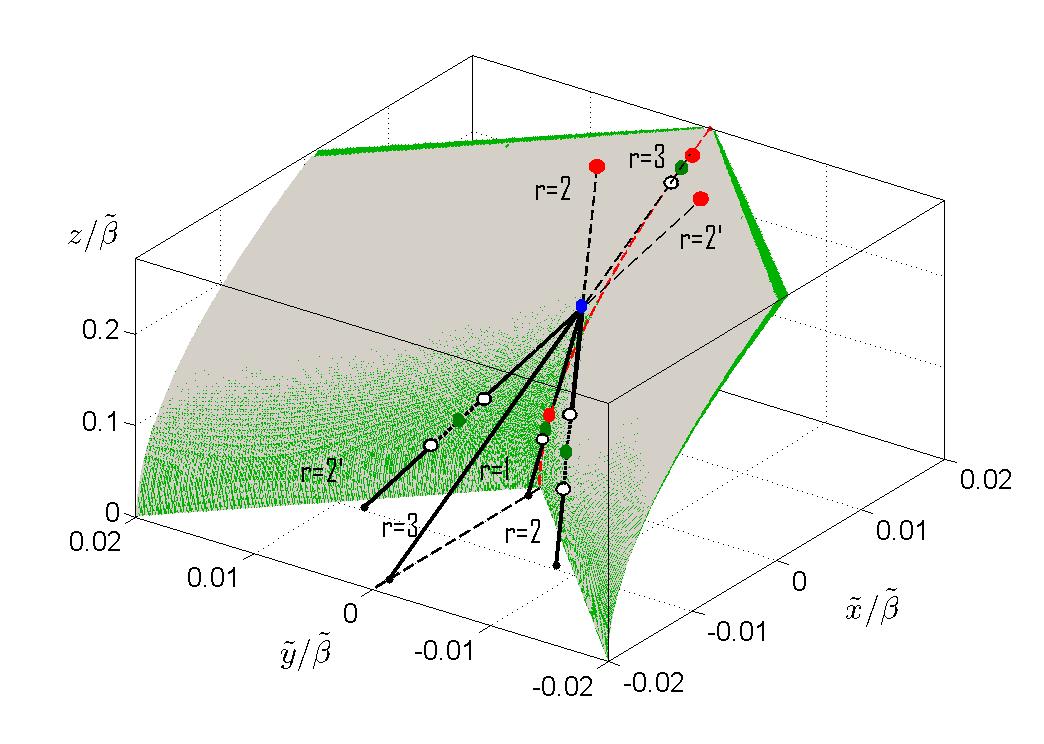}
\caption { (Color online)  Rays  reaching an observation point at $\3r=(\tilde x,\tilde y,z)=(0.005,0,0.17)\tilde\beta$ (blue point) on the lit side of the caustic, tagged by the ray index $r=1,...4$. The parts of the rays before and after $\3r$ are depicted as solid and dashed lines, respectively, and the parts between caustic 1 and 2 are depicted as dotted lines. The ray initiation points $(\tilde x'_r,\tilde y'_r)/\tilde\beta$ are: ray~1 $(-0.14,0)10^{-2}$; rays~2,3 $(-0.95,\mp.81)10^{-2}$; ray~4 $(-1.8,0)10^{-2}$. The ray direction parameters $(\xi_r,\eta_r)$ are: ray~1 $(2.7,2.7)10^{-2}$; ray~2 $(2.7,9.4)10^{-2}$; ray~3 $(9.4,2.7)10^{-2}$; ray~4 $(9.5,9.5)10^{-2}$.
The paths of rays~2 and ~3 are similar to that of ray $s=1$ in Fig.~\ref{Fig_caustic1_ray_skeleton}: they penetrate caustic~1 (through the white circles), touch caustic~2 (green points), penetrate caustic~1 again (upper white circles) and then touch caustic~1 (red points) \emph{after} passing through $\3r$   (blue point). Rays~1 and~4 propagate in the $\tilde y=0$ plane. Ray~4  touches the caustic \emph{after} passing through $\3r$: It penetrates caustic~1 (through the white circle),  touches the edge of caustic~2 (green point) and then penetrates back through the cusped-edge of caustic~1 and at the same point it is also tangent to the cusp (red point; see discussion in the last paragraph of Sec.~\ref{caustic_top_parax_exact} and also Fig.~\ref{Fig_caustic_coord}). The path of ray~1 is similar to that of ray~4, only it already touched the caustics \emph{before} passing through $\3r$. When $\3r$ moves toward the cusped-edge of caustic~1 from its lit side, rays~1--3 coalesce. When $\3r$ is on the shadow side of edges~1 such that it is between the edges of caustics~1 and~2, it is  reached by 2 rays which penetrate caustic 1, one which results from the coalesced rays~1-3 and has already touched caustic~2, and ray~4 which has not touched caustic~2 yet. These two rays coalesce as $\3r$ moves toward the edge of caustic~2. }
 \label{Fig_rays}
 \end{figure}

\begin{figure}
\centering
\includegraphics[width=8cm]{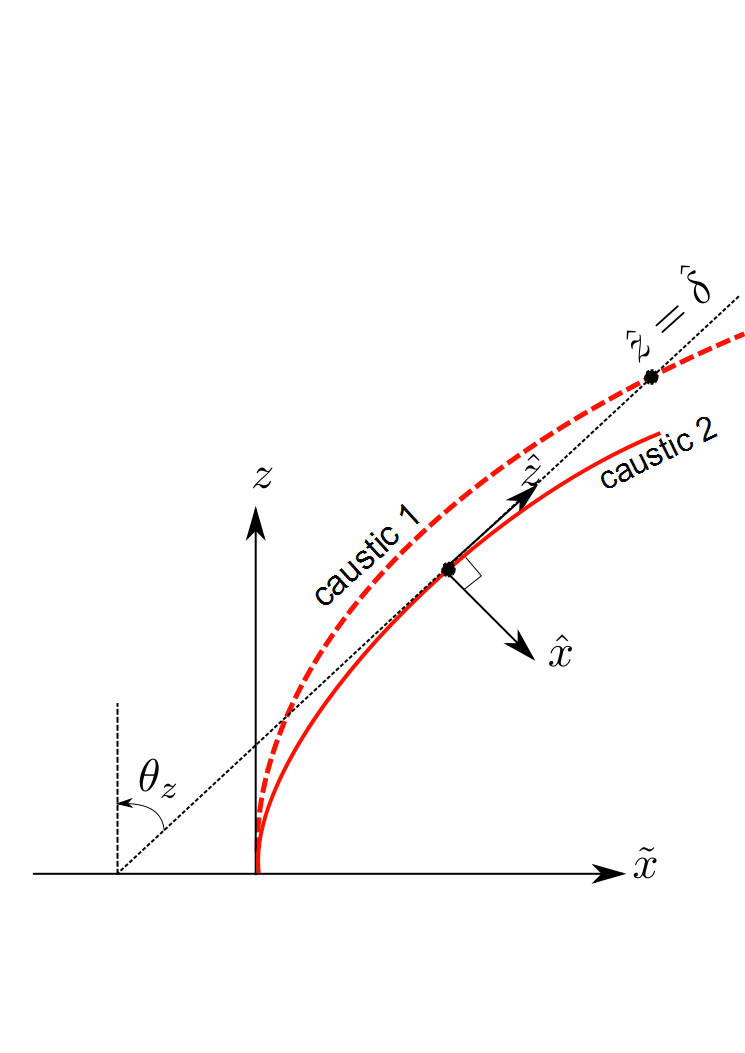}
\caption {(Color online)  Schematic view of the edges of caustics~1 and~2 (dashed and solid red lines, respectively) in the symmetry plane $\tilde y=0$. The edges are  given in Eqs.~\eqref{edge1_eq}--\eqref{edge2_eq}. As shown in Fig.~\ref{Fig_caustic_cross_z}, caustic 1 is actually a cuspoid but only its edge is shown  here.  A ray (dashed black line) penetrates through caustic~1 before the point of tangency with caustic~2. After that point it penetrates again through caustic~1 (point $\hat z=\hat \delta$ in the figure). At that point it is also tangent to the cusp of caustic~1 (see see discussion in the last paragraph of Sec.~\ref{caustic_top_parax_exact}).  Also shown, the local Cartesian coordinate system $(\hat x,\hat y,\hat z)$ centered around a point on  caustic~2, such that $\hat z$ is the tangent to the edge at that point, while $\hat x$ and $\hat y=\tilde y$ are the normal and the bi-normal there, respectively. This system is parameterized by the angle $\theta_z$ of the $\hat z$-axis with respect to $z$-axis, and it is used to construct the canonical integral of the field in Sec.~\ref{catastrophe}. }
 \label{Fig_caustic_coord}
\end{figure}

\begin{figure}
\centering
 \subfigure[$\bar y=0$]{%
      \label{Fig_cat_field_a}%
      \includegraphics[width=8cm]{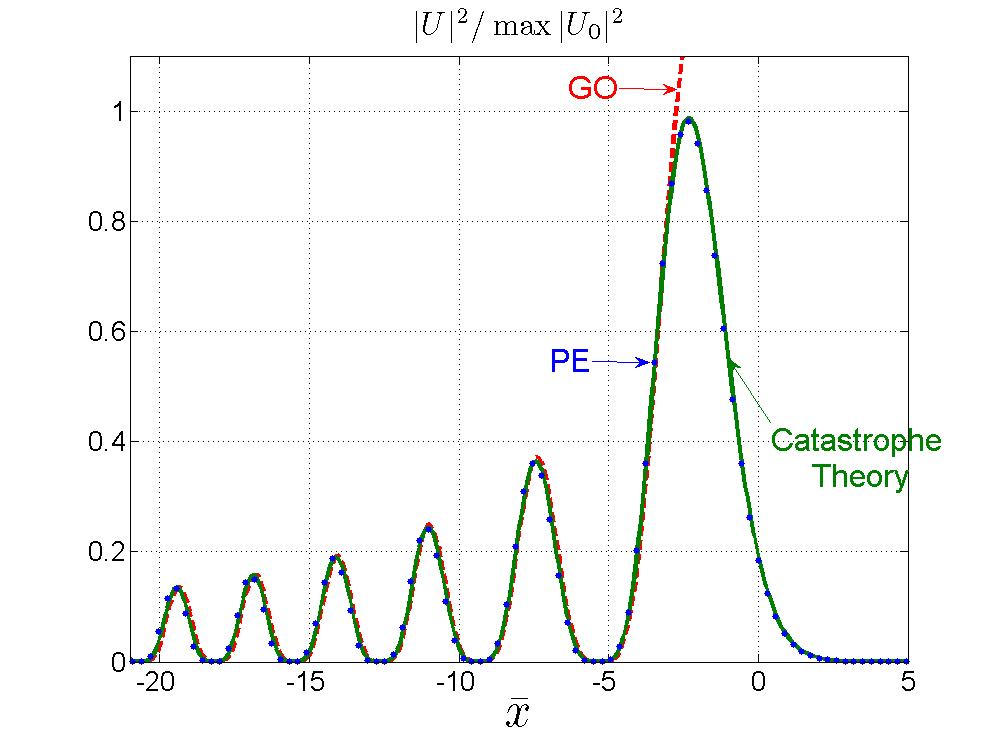}}%
   \subfigure[$\bar y=1$]{%
    \label{Fig_cat_field_b}%
    \includegraphics[width=8cm]{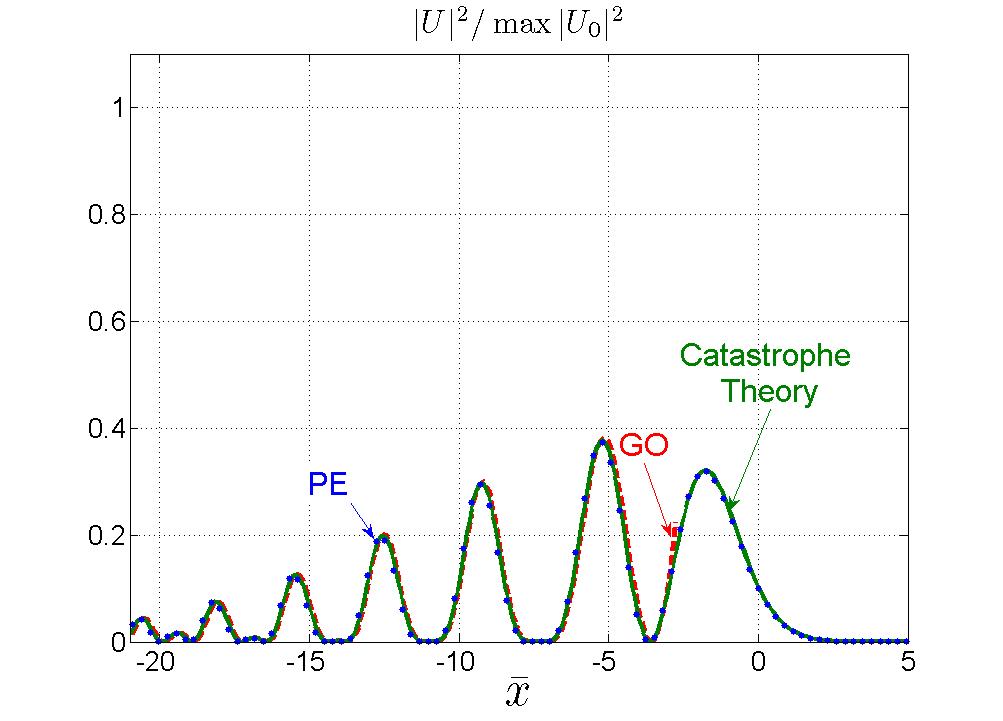}}%
\caption {(Color online) Intensity $|U|^2$ of the field in the planes (a) $\bar y=0$ and (b) $\bar y=1$, plotted along $\bar x$, the normalized $\hat x$ axis of Fig.~\ref{Fig_caustic_coord}, which is normal to the beam axis, and centered at a point on this axis; here this point is $(\tilde x,z)= \tilde\beta(0.006, 0.16)$.  The fields were calculated via the catastrophe theory (green line), the PE approximation (blue dots) and GO (red line). The intensity is normalized with respect to $\max |U_\70|^2=\9{Ai}^2(0)$ of  Eq.~\eqref{aperture_field}. Parameters: $k\tilde\beta=10^4$, $\tilde \alpha=10^{-4}$.}
\label{Fig_cat_field}
\end{figure}

\begin{figure}
\centering
 \subfigure[$\bar y=0$]{%
      \label{Fig_cat_field_c}%
      \includegraphics[width=8cm]{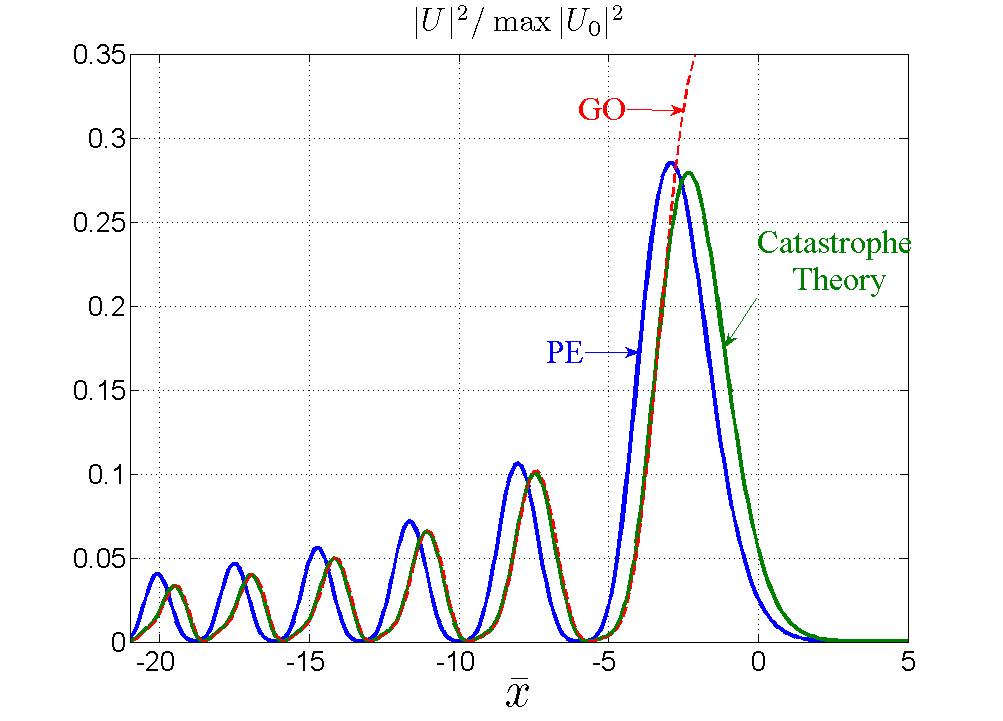}}%
   \subfigure[$\bar y=1$]{%
    \label{Fig_cat_field_d}%
    \includegraphics[width=8cm]{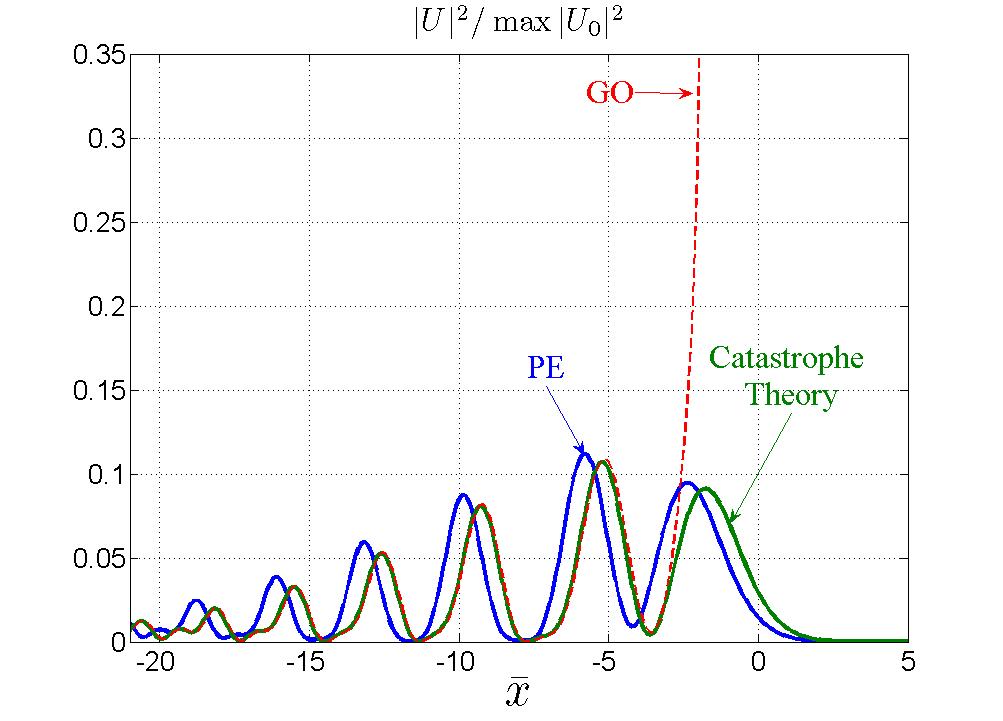}}%
\caption {(Color online)  Same as Fig.~\ref{Fig_cat_field} but for $k\beta=10^6$. }
\label{Fig_cat_field2}
\end{figure}

\begin{figure}[tb]
\centering
      \includegraphics[width=11cm]{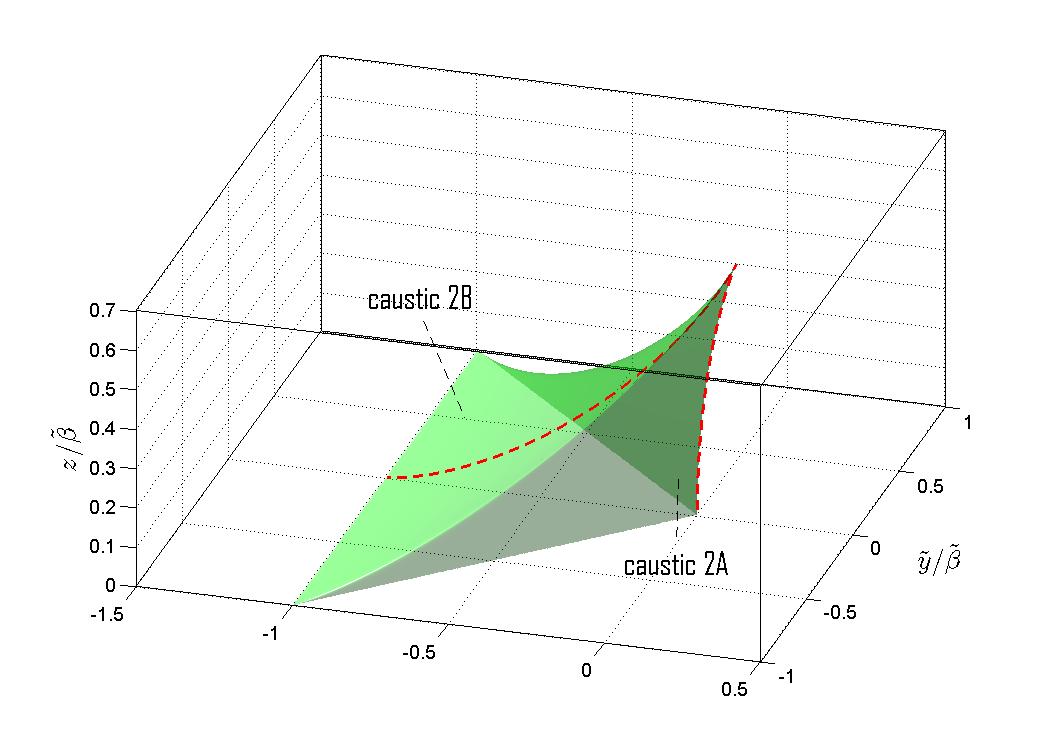}
\caption {(Color online)  The global structure of caustic~2 (green surface). ``Caustic~2A'' is the part that was shown in Figs.~\ref{Fig_caustic_exact} and~\ref{Fig_caustic2_ray_skeleton}, whose edge follows the beam axis. This part of the caustic is formed by rays leaving the aperture at points close to the caustic. ``Caustic~2B'' is an additional part of caustic~2 not shown in Fig.~\ref{Fig_caustic_exact} and it is formed by rays leaving the aperture sideways from faraway points (see Fig.~\ref{Fig_caustic_cross_sym}).  At some range, caustic~2  terminates at a pyramid-like edge. Red dashed-line: cross section of the caustic at $\tilde y=0$ which is also shown as red dashed-line in  Fig.~\ref{Fig_caustic_cross_sym}.}
 \label{Fig_caustic3}
 \end{figure}

\begin{figure}[tb]
\centering
      \subfigure[The part of caustic~1 formed by species~1]{%
      \label{Fig_caustic4}%
      \includegraphics[width=11cm]{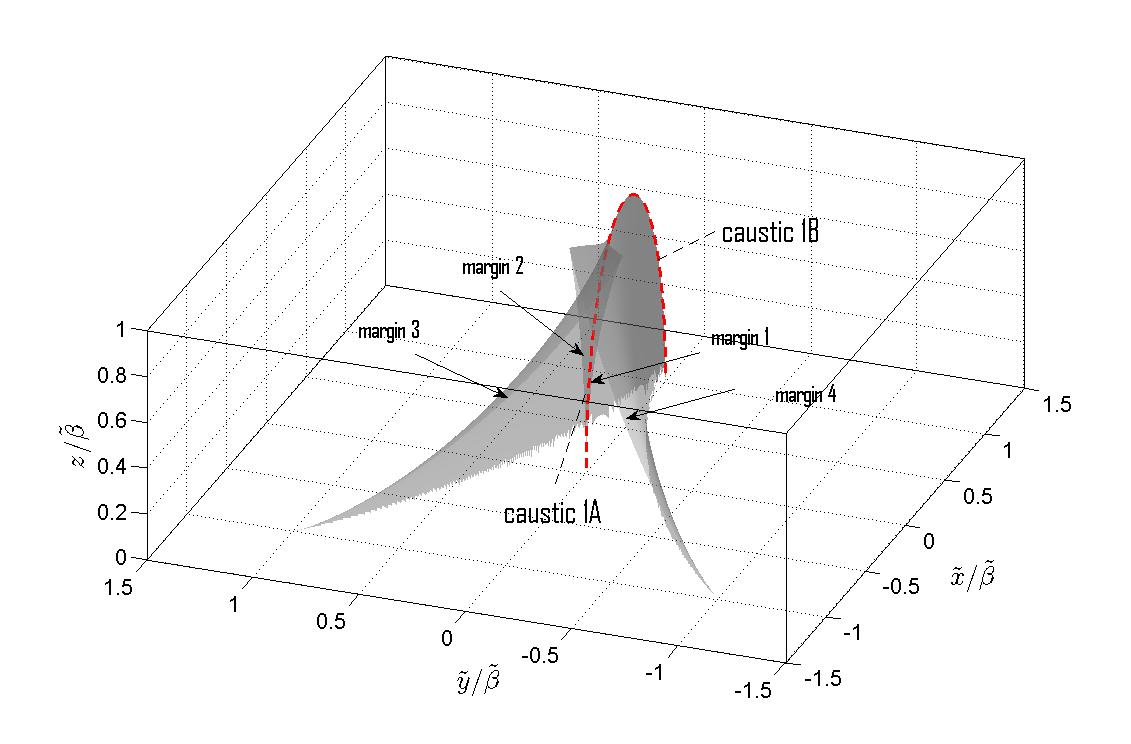}}%
       \\
       \subfigure[The part of caustic~1 formed by species~2,3,4]{%
      \label{Fig_caustic5}%
      \includegraphics[width=10.5cm]{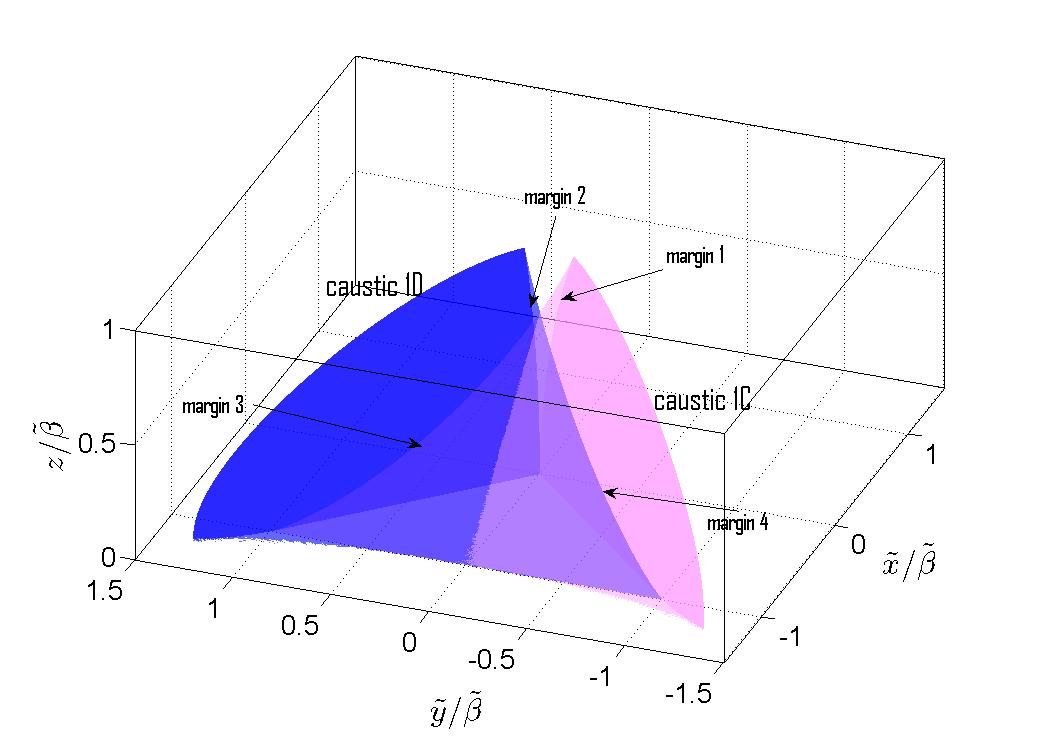}}%
\caption {(Color online)  The global structure of  caustic~1. Since the structure is complex, we show separately the parts formed by ray species $s=1$ and by species $s=2,3,4$ (Figs.~(a) and~(b), respectively). In~(a), caustic~1A is the caspoid shown in Figs.~\ref{Fig_caustic_exact} and~\ref{Fig_caustic1_ray_skeleton} which is formed by rays leaving the aperture at points near the beam axis, while caustic~1B is formed by rays leaving the aperture sideways at faraway points. In~(b), caustic~1C (blue) and caustic~1D (magenta) are two overlapping sheets connect to caustic~1B at margins~3 and~4 and to caustic~1A  at margins~1 and~2 (see also  cross-sectional cuts of the caustic in Fig.~\ref{Fig_cross_hat}).  }
 \label{Fig_caustic45}
 \end{figure}

 \begin{figure}[tb]
\centering
      \subfigure[$z/\tilde\beta=0.2$.]{%
      \label{Fig_a}%
      \includegraphics[height=5.5cm]{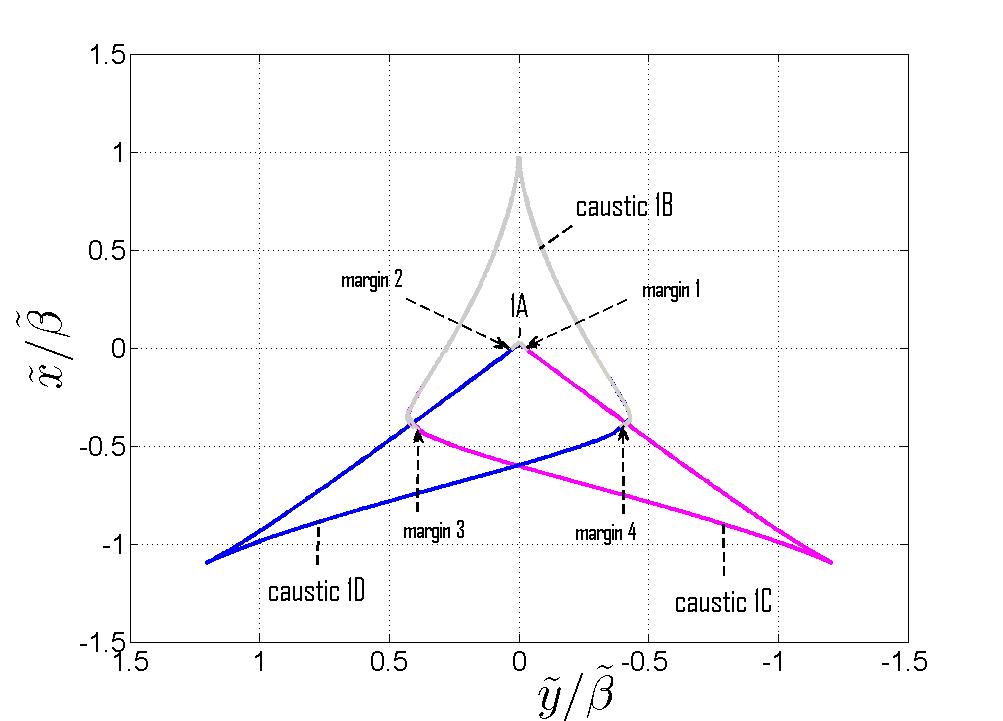}}%
       \subfigure[$z/\tilde\beta=0.4$.]{%
      \label{Fig_b}%
      \includegraphics[height=5.5cm]{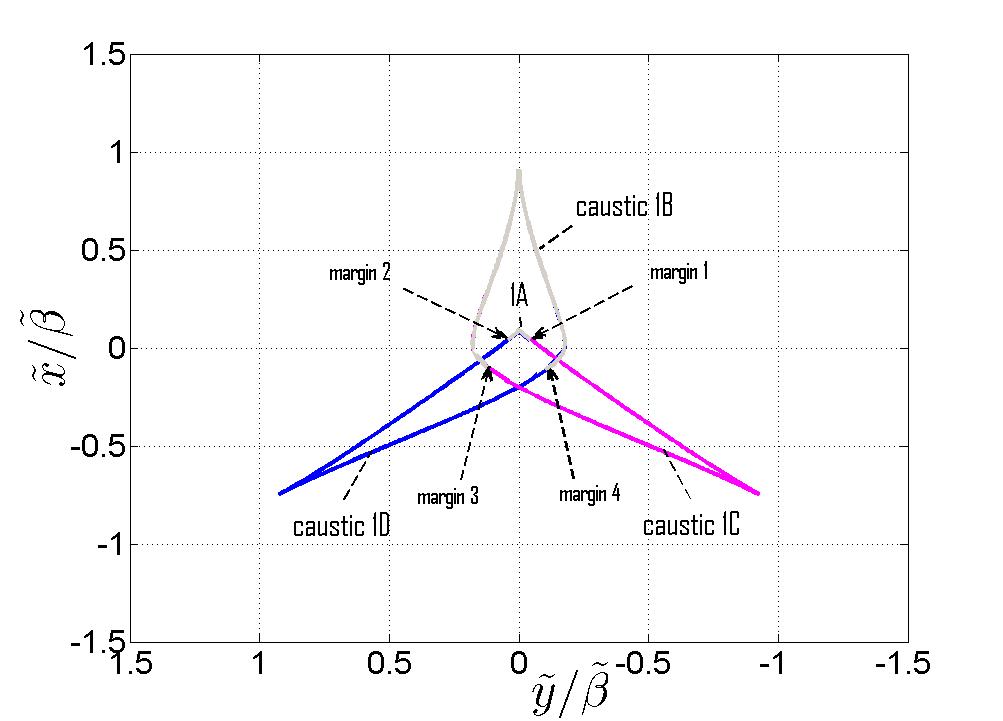}}%
\caption {(Color online)  Cross sections of caustic~1 in Fig.~\ref{Fig_caustic45} at $z$ constant planes.}
 \label{Fig_cross_hat}
 \end{figure}

 \begin{figure}[tb]
\centering
      \includegraphics[width=10cm]{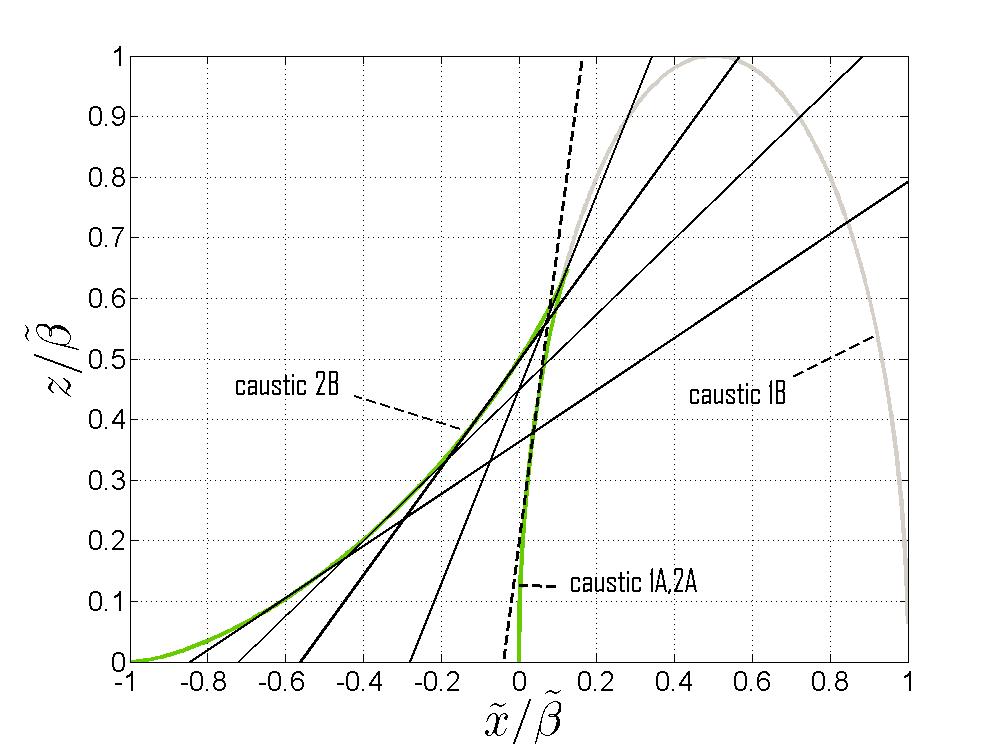}
\caption {(Color online) Complete cross section of caustic~1 and~2 (gray and green lines, respectively) at the plane of symmetry $\tilde y=0$, also shown as red dashed-lines in Figs.~\ref{Fig_caustic3} and \ref{Fig_caustic45}.  These lines are defined in Eqs.~\eqref{edge1_coord}--\eqref{edge2_coord} (or alternatively by Eqs.~\eqref{edge1_eq}--\eqref{edge2_eq}). Black solid lines: rays emanating sideways from faraway points, forming caustics~1B and~2B.  Black dashed line: A typical ray near the axis forming caustic~1A and~2A (indistinguishable within the scale of the figure). Caustics~1C and~1D of Fig.~\ref{Fig_caustic5} are not formed by rays contained in the $\tilde y=0$ plane and are therefore omitted here.}
 \label{Fig_caustic_cross_sym}
 \end{figure}

\end{document}